\documentclass[a4paper, 12pt, titlepage]{article}

\usepackage[a4paper,top=2.5cm,bottom=2.5cm,left=3.3cm,right=2.2cm]{geometry}

\usepackage{fancyhdr}
\usepackage[turkish,american]{babel}
\usepackage[T1]{fontenc}
\usepackage{microtype}

\usepackage[utf8]{inputenc}

\usepackage{subcaption}

\usepackage{float}

\usepackage{longtable}
\usepackage{array}

\usepackage{enumitem}

\usepackage{helvet}
\usepackage{sectsty}
\allsectionsfont{\normalfont\sffamily\bfseries}
\sectionfont{\fontsize{18pt}{21.6pt}\sffamily\bfseries}
\subsectionfont{\fontsize{16pt}{19.2pt}\sffamily\bfseries}
\subsubsectionfont{\fontsize{14pt}{16.8pt}\sffamily\bfseries}

\usepackage{courier}

\usepackage{tocloft}

\tocloftpagestyle{fancy}

\usepackage{hyperref}
\hypersetup{
    unicode=true,
    colorlinks=true,
    urlcolor=blue,
    citecolor=black,
    menucolor=black,
    linkcolor=black
}

\usepackage[labelfont=bf,font=sf]{caption}

\usepackage{algorithmicx}
\usepackage{algpseudocode}
\usepackage{float}
\usepackage{makecell}
\usepackage{tabularx}
\newcolumntype{Y}{>{\centering\arraybackslash}X}

\usepackage{listings}
\usepackage{amsmath}
\usepackage{caption}
\usepackage[most]{tcolorbox}

\newcommand{\thetitle}{StajChain: A Hyperledger Fabric-Based
Multi-Party Internship Agreement System}
\newcommand{\theturkishtitle}{StajChain: Hyperledger Fabric Tabanlı Çok Paydaşlı Staj Sözleşmesi Sistemi}
\newcommand{\theauthor}{Rampia Perente}
\newcommand{\thedate}{May 2026}
\newcommand{\theturkishdate}{Mayıs 2026}

\usepackage{color}
\definecolor{darkgreen}{RGB}{0, 128, 0}

\newcommand\Includegraphics{\expandafter\includegraphics\expandafter} 

\title{\thetitle}
\author{\theauthor}
\date{\thedate}

\usepackage{graphicx}

\begin{document}
\numberwithin{figure}{section}
\numberwithin{table}{section}
\numberwithin{lstlisting}{section}
\shorthandoff{=}

\begin{titlepage}
    \bfseries 
    \sffamily 
    \begin{center}
        \LARGE{\textbf{ISTANBUL TECHNICAL UNIVERSITY \\ 
               FACULTY OF COMPUTER AND INFORMATICS} } \\
        \vspace{5.5cm}
        \LARGE{\thetitle}  \\
        \vspace{4.5cm}
        \Large{Graduation Project Final Report} \\
        \vspace{0.5cm}
        \Large{\theauthor} \\
        \Large{150220916} \\
        \vspace{4cm}
        \large{Department: Computer Engineering} \\
        \vspace{1.5cm}
        \large{Advisor: Prof. Dr. Şerif Bahtiyar} \\
        \vspace{\fill} 
        \large{\normalfont \sffamily \thedate}
    \end{center}
\end{titlepage}

\pagenumbering{Roman} 
\newpage
\section*{Statement of Authenticity}
I/we hereby declare that in this study
\begin{enumerate}
    \item all the content influenced from external references are cited clearly and in detail,
    \item and all the remaining sections, especially the theoretical studies and implemented software/hardware that constitute the fundamental essence of this study is originated by my/our individual authenticity.
\end{enumerate}
\vspace{1em}
Istanbul, \theturkishdate
\vspace{3em}\\ \theauthor

\newpage
\section*{Acknowledgments}
I am grateful to my supervisor, Prof. Dr. Şerif Bahtiyar, for his guidance and support throughout this project.

I would also like to thank my family for their endless support and efforts from the beginning of my educational journey to the present.

\newpage
\section*{\centering\thetitle}
\centerline{\fontsize{16pt}{21.6pt}\sffamily\bfseries (SUMMARY)}

StajChain is a permissioned blockchain-based system that manages multi-party internship agreement processes. The main motivation of this project is dependency on paper-based documents, handwritten signatures, manual verification steps, and limited traceability in internship approval workflows. Different parties, such as students, companies, faculties, and university internship offices, take part in these workflows. Many interactions and document exchanges between these parties mostly happen outside a controlled system. Thus, documents can be modified, lost, or approved without proper verification. Also, tracking the agreement approval history becomes difficult.

The StajChain project proposes a Hyperledger Fabric-based solution for the defined problems. Hyperledger Fabric is selected because the system requires known and authorized participants, controlled access, and role-based permissions. The main actors are students, companies, faculty users, and the central internship unit. Each of these actors has a separate role and can perform only the allowed actions. The system supports agreement creation, student approval, company approval, faculty approval, central activation, completion, rejection, and agreement history queries.

For the implementation, the system is built with a React frontend, a Node.js/Express backend, an SQLite off-chain database, and a Hyperledger Fabric blockchain layer. The backend handles authentication, role-based access control, business rule validation, database operations, and Fabric transaction invocation. Meanwhile, SQLite stores application-level data such as users, companies, company requests, and agreement metadata. On the other hand, the Hyperledger Fabric ledger stores the core internship agreement records, agreement states, and lifecycle state transitions. 

Furthermore, smart contracts in the ledger layer are used to control the agreement workflow. They check the caller role, actor identity, current agreement status, and allowed state transitions before updating the ledger. This ensures that each approval step is performed in the allowed order and any user who invokes a transaction is authorized. Other internship-specific validations are divided between the backend and the smart contract. The backend checks application-level constraints such as minimum credit requirements, internship start-date constraints, company availability, and student metadata consistency before sending a transaction. The smart contract, on the other hand, enforces ledger-side rules such as caller authorization, allowed state transitions, schedule consistency, overlap prevention, and agreement history access.

The proposed system is evaluated using different functional and performance tests. The functional tests evaluated the complete agreement lifecycle, agreement creation rules, business-rule enforcement, rejection handling, company registration workflow, invalid inputs, unauthorized actions, and incorrect workflow transitions. The results showed that valid operations were completed successfully and invalid operations were rejected as expected. 

Moreover, the performance tests showed that Fabric-based agreement read operations have a Transactions Per Second (TPS) value between 124.52 and 159.53 and a P95 latency below 100 ms. Meanwhile, ledger write operations had approximately 2 seconds P95 latency since Fabric write transactions require endorsement, ordering, validation, and commit processing before updating the ledger state. These results are acceptable since StajChain is an internship agreement workflow where correctness, traceability, and controlled access are more important than high-frequency transaction processing.

In summary, this project introduces a secure and transparent internship agreement management system based on permissioned blockchain technology. Smart contracts, role-based access control, and distributed ledger technology are combined to provide a digital and reliable process, addressing the limitations in traditional paper-based internship workflows. As future work, the prototype can be deployed on a distributed multi-organization Hyperledger Fabric network. Also, secure document storage, improved company verification, notification mechanisms, and reporting tools can be added.

\newpage
\section*{\centering\theturkishtitle}
\centerline{\fontsize{16pt}{21.6pt}\sffamily\bfseries (ÖZET)}

StajChain, çok taraflı staj sözleşmesi süreçlerini yöneten izinli blok zinciri tabanlı bir sistemdir. Bu projenin temel motivasyonu, staj onay iş akışlarında kağıt tabanlı belgelere, el yazısı imzalara, manuel doğrulama adımlarına ve sınırlı izlenebilirliğe olan bağımlılıktır. Öğrenciler, şirketler, fakülteler ve merkezi staj birimi gibi farklı taraflar bu iş akışlarına katılır. Bu taraflar arasındaki birçok etkileşim ve belge alışverişi çoğunlukla kontrollü bir sistemin dışında gerçekleşir. Bu nedenle belgeler değiştirilebilir, kaybolabilir veya uygun doğrulama yapılmadan onaylanabilir. Ayrıca, sözleşme onay geçmişinin izlenmesi zorlaşır.

StajChain projesi, tanımlanan sorunlar için Hyperledger Fabric tabanlı bir çözüm önermektedir. Hyperledger Fabric, sistemin bilinen ve yetkilendirilmiş katılımcılar, kontrollü erişim ve rol tabanlı izinler gerektirmesi nedeniyle seçilmiştir. Ana aktörler öğrenciler, şirketler, fakülte kullanıcıları ve üniversite merkezi staj birimidir. Bu aktörlerin her birinin ayrı bir rolü vardır ve yalnızca izin verilen eylemleri gerçekleştirebilir. Sistem, sözleşme oluşturma, öğrenci onayı, şirket onayı, fakülte onayı, merkezi aktivasyon, tamamlama, reddetme ve sözleşme geçmişi sorgulamalarını destekler.

Uygulama için sistem, React ön yüzü, Node.js/Express arka yüzü, SQLite zincir dışı veritabanı ve Hyperledger Fabric blok zinciri katmanı ile oluşturulmuştur. Arka yüz, kimlik doğrulama, rol tabanlı erişim kontrolü, iş kuralı doğrulaması, veritabanı işlemleri ve Fabric ağı ile iletişimi yönetir. SQLite ise kullanıcılar, şirketler, şirket talepleri ve sözleşme meta verileri gibi uygulama düzeyindeki verileri depolar. Öte yandan, Hyperledger Fabric defteri, temel staj sözleşmesi kayıtlarını, sözleşme durumlarını ve yaşam döngüsü durum geçişlerini depolar.

Ayrıca, blok zinciri katmanındaki akıllı sözleşmeler, sözleşme iş akışını kontrol etmek için kullanılır. Akıllı sözleşmeler, ledger durumunu güncellemeden önce arayanın rolünü, aktör kimliğini, mevcut sözleşme durumunu ve izin verilen durum geçişlerini kontrol eder. Bu, her onay adımının doğru sırada gerçekleştirilmesini ve işlemi çağıran kullanıcının yetkili olmasını sağlar. Sistem ayrıca minimum kredi gereksinimleri, staj başlangıç tarihi kısıtlamaları, çalışma günü doğrulaması, çakışma önleme ve şirketin sistemde kayıtlı ve onaylı olması gibi staja özgü kuralları da kontrol eder.

Önerilen sistem, farklı fonksiyonel ve performans testleri kullanılarak değerlendirilmiştir. Fonksiyonel testler, sözleşme yaşam döngüsünün tamamını, sözleşme oluşturma kurallarını, iş kuralı uygulamasını, ret işlemlerini, şirket kayıt iş akışını, geçersiz girdileri, yetkisiz işlemleri ve hatalı iş akışı geçişlerini değerlendirmiştir. Sonuçlar, geçerli işlemlerin başarıyla tamamlandığını ve geçersiz işlemlerin beklendiği gibi reddedildiğini göstermiştir.

Ayrıca, performans testleri, Fabric tabanlı sözleşme okuma işlemlerinin Saniye Başına İşlem (TPS) değerinin 124,52 ile 159,53 arasında ve P95 gecikmesinin 100 ms'nin altında olduğunu göstermiştir. Bu arada, defter yazma işlemlerinin P95 gecikmesi yaklaşık 2 saniyedir, çünkü Fabric yazma işlemleri ledger durumunu güncellemeden önce onaylama, sıralama, doğrulama ve commit işlemlerini gerektirir. Bu sonuçlar kabul edilebilir, çünkü StajChain yüksek sıklıkta işlem gerçekleştirmeden ziyade doğruluk, izlenebilirlik ve kontrollü erişimin daha önemli olduğu bir staj sözleşmesi iş akışıdır.

Özetle, bu proje izinli blok zinciri teknolojisine dayalı güvenli ve şeffaf bir staj sözleşmesi yönetim sistemi sunmaktadır. Akıllı sözleşmeler, rol tabanlı erişim kontrolü ve ledger geçmişi birlikte kullanılarak geleneksel kağıt tabanlı staj iş akışlarındaki sınırlamaları azaltan dijital ve izlenebilir bir süreç sağlanmıştır. Gelecekteki çalışmalarda, prototip dağıtılmış çok kuruluşlu bir Hyperledger Fabric ağına konuşlandırılabilir. Ayrıca, güvenli belge depolama, geliştirilmiş şirket doğrulama, bildirim mekanizmaları ve raporlama araçları da eklenebilir.

\newpage
\tableofcontents
\newpage

\pagenumbering{arabic}
\section{Introduction and Project Summary}
Despite advancements in technology, many administrative processes are still conducted using paper-based documents and require handwritten signatures and manual verification steps. Approval workflows involving multiple stakeholders often rely on physical forms or scanned copies that must be manually shared, checked, and approved. Internship agreement processes are a typical example of such workflows because they involve students, companies, faculty units, and central university offices. These processes require not only document exchange, but also trust, traceability, and correct execution of approval steps.

In internship agreement processes, documents typically require approval from more than one party before the internship starts. Each party is expected to approve the agreement in the correct order. Also, the final document should remain verifiable after these approvals. However, when signed documents are exchanged outside a controlled digital system, several problems may occur. Documents can be lost, modified, approved without proper verification, or submitted in different versions. Thus, ensuring document integrity and tracking the approval history becomes challenging.

On the other hand, blockchain technology has been widely used in systems that require immutable records, transparent history, and controlled process execution. Several studies have used blockchain-based approaches for document management, administrative workflows, business workflow execution, and educational records \cite{das2022management, azzam2023OCRblockchain, rahat2025eviblock, krogsboll2020smart, belhi2021odoo, pairetti2023collaborative, leporati2024certification, turkanovic2018eductx, han2018educationrecords, grather2018lifelong, dash2025hyperledger, saleh2020certificates}. These studies indicate that blockchain is suitable for processes such as internship agreements, where multiple parties need to approve and verify records without depending on central manual control.

The main focus of this project is the internship agreement and management processes at Istanbul Technical University (ITU). In the current workflow, students start the internship record process through the university system and generate an agreement document after entering the necessary information. Then, the student and the company sign this document consecutively. Later, the document is uploaded back to the system, reviewed by the faculty internship commission, and forwarded to the central internship unit for insurance procedures. Although the faculty approval and the process after that are done inside the university system, the exchange and approval of documents between the student and the company take place outside the university system. This creates risks related to document manipulation, forged approvals, and limited traceability.

For these reasons, internship approval processes require systems that support digitally verifiable agreement workflows. The StajChain project proposes a permissioned blockchain-based internship agreement management system. Since actors in the system must be known and authorized and must have controlled access rights, a permissioned blockchain is preferred. Permissioned blockchain systems are more suitable for institutional workflows than public blockchain networks \cite{vukolic2016quest}. In the StajChain project, Hyperledger Fabric is used as the blockchain platform since it supports controlled multi-organizational workflows and chaincode to implement application-specific rules \cite{androulaki2018fabric, cachin2016architecture, li2020survey}. To control the internship agreement lifecycle, smart contracts are used. Also, all critical agreement data and state changes are stored immutably on the ledger in the blockchain layer. 

In the application layer, the backend is developed with Node.js and manages authentication, role-based access, company registration, and communication with the blockchain network. SQLite is used as the database to store off-chain application data, while the frontend provides React interfaces for students, companies, faculty users, and the central internship unit.

The proposed system digitalizes the internship agreement lifecycle. Each entity in the system, such as students, companies, faculty users, and the central internship unit, has separate roles. Also, the system follows a predefined sequence of steps, as each actor can perform specific actions, such as approvals and rejections, only at the appropriate stage of the workflow. Moreover, the system supports history tracking and controlled company registration, where the central internship unit reviews company requests instead of adding them directly to the system. This ensures that each actor in the prototype is known and authorized.

Overall, the main contributions of this project are as follows:

\begin{itemize}
    \item It proposes a permissioned blockchain-based system for more secure, verifiable, and traceable internship agreement management.
    
    \item It implements a smart contract-based workflow to control the sequential agreement steps and allows querying ledger history by storing all state changes and critical data on-chain.
    
    \item It implements a working prototype using Hyperledger Fabric and Node.js, which can be used in many different systems in the future.
\end{itemize}

All in all, this project aims to provide a transparent, verifiable, and secure internship agreement lifecycle for university internship processes while eliminating paper-based agreement workflows. This implementation is a prototype and demonstrates how a permissioned blockchain can be used to manage agreement-based administrative workflows involving multiple stakeholders with different roles and responsibilities.

\newpage

\section{Comparative Literature Survey}
\subsection{Blockchain Background}
Blockchain technology is based on earlier studies about timestamping, chained records, and decentralized trust. Initially, Haber and Stornetta proposed a method for securing digital documents by using timestamps. This was a hash-based document linking that allowed detection of any modification \cite{haber1991timestamp}. Bayer et al. improved this idea with a more efficient timestamping structure \cite{bayer1993improving}. Later, Nakamoto introduced Bitcoin as a peer-to-peer electronic cash system. He showed that a distributed ledger can store records in an immutable chain without a central authority \cite{nakamoto2008bitcoin}. Smart contracts extended this idea by allowing the automatic execution of predefined rules on blockchain networks \cite{szabo1996smart}. These studies became the theoretical background of the StajChain project, which also uses immutable ledger records and smart contracts to manage the internship agreements.

\subsection{The Use of Blockchain in Document and Education Workflows}
Blockchain technology has been used in many domains where data integrity, traceability, and controlled execution are required. This usage inspired the StajChain project, which manages internship agreements that pass through multiple approval steps, and where the integrity of agreement records and the visibility of previous actions are important. Similar requirements are discussed in multiple studies that focus on blockchain-based administrative workflows and education systems. 

\subsubsection{Document and Administrative Workflow Systems}
Several studies address secure record keeping and workflow traceability by applying blockchain to document management and administrative workflow control. To exemplify, Das et al. proposed a blockchain-based document management framework for construction applications. They used smart contracts for irreversible approval workflow logic, ledger records for document changes, and a blockchain-based structure for version history \cite{das2022management}. Moreover, Azzam et al. introduced SECHash to manage incoming documents in e-government by combining blockchain and Optical Character Recognition (OCR) to protect accepted documents against modification or loss \cite{azzam2023OCRblockchain}. Also, Evi-Block is proposed, which is a blockchain-based evidence management system for secure and auditable handling of digital evidence records \cite{rahat2025eviblock}.

Furthermore, Krogsbøll et al. used smart contracts for a municipal government process and showed that administrative steps can be controlled through blockchain \cite{krogsboll2020smart}. Other studies also used blockchain and smart contracts for administrative and enterprise workflows, such as Odoo workflow management, collaborative business process execution, and workflow certification \cite{belhi2021odoo, pairetti2023collaborative, leporati2024certification}. The StajChain project does not just focus on the integrity or transparency of agreement records. It also controls the correct execution of the internship agreement approval lifecycle between multiple parties using smart contracts.

\subsubsection{Education and Academic Certificate Systems}
Another common application area of blockchain is academic certificate verification and storage systems due to the need for trusted issuance, controlled access, and verifiability among multiple parties. For instance, Turkanović et al. introduced EduCTX, which is a higher education credit platform that supports the management and transfer of academic credit records among higher education institutions, students, and other stakeholders such as companies \cite{turkanovic2018eductx}. Furthermore, Han et al. proposed a blockchain-based education record verification system where people can keep their education records and share them with employers or others \cite{han2018educationrecords}. Also, Gräther et al. implemented the Blockchain for Education platform, which issues, validates, and shares certificates between certification authorities, learners, and employers \cite{grather2018lifelong}. 

Other studies also focused on certificate authentication and educational credential management. Dash et al. developed a secure academic certificate authentication system to reduce certificate fraud and enable verification by universities, students, and companies \cite{dash2025hyperledger}. Similar approaches were also introduced by Saleh et al. \cite{saleh2020certificates}. In addition, Jirgensons and Kapenieks discussed the potential of blockchain-based digital learning credentials to ensure permanent records, enable direct user access, and reduce administrative workload for universities \cite{jirgensons2018credential}.

These studies show that blockchain is widely used in education for certificate verification, academic record sharing, credit transfer, and lifelong learning credentials. However, most of them focus on issuing, storing, sharing, or verifying already-created academic records. In contrast, StajChain focuses on the internship agreement process itself, where an agreement is created, reviewed, approved, rejected, activated, completed, and traced through multiple role-based steps.

\subsection{Hyperledger Fabric and Performance Studies}
The studies discussed above show that blockchain can support document management, administrative workflows, and education records. Since blockchain networks differ in access control, participant identity, and performance characteristics, the choice of blockchain network type becomes important for such systems.

\subsubsection{Permissioned Blockchain and Hyperledger Fabric}

In blockchain systems, there is a classification of networks as public, private, and consortium according to their permissioning and access rules \cite{feng2019survey}. These terms define who can access the blockchain network, submit transactions, and participate in the consensus process. Public blockchains such as Bitcoin and Ethereum are open to all users, and they can view blockchain records and submit transactions without permission from a central authority \cite{feng2019survey}. However, institutional applications usually require known users, controlled access, and role-based permissions. In addition, Proof of Work (PoW), which is the mining-based consensus mechanism used in many early public blockchains, requires high computational resources, time, and electricity for consensus \cite{guo2022survey}. This makes PoW-based public blockchains less suitable for administrative workflows. 

For these reasons, permissioned blockchain platforms are more suitable for systems where participants have known identities and their access is controlled. In these systems, the consensus mechanism is performed among a known set of nodes instead of anonymous mining as in PoW. Byzantine Fault Tolerance (BFT) is an important approach for such systems because it allows distributed nodes to reach agreement even if some nodes behave incorrectly or maliciously \cite{castro2002practical}. Vukolić compared PoW-based blockchains with BFT-based approaches and showed that PoW-based blockchains have high latency and limited throughput, while BFT-based approaches are more suitable for smaller networks with known node identities \cite{vukolic2016quest}. Later, Cachin and Vukolić analyzed different consensus approaches and blockchain platforms and argued that permissioned blockchains achieve higher performance compared to public ones since network participants are known and authorized \cite{cachin2017consensus}. These studies guided the selection of a blockchain platform for the StajChain project because the system includes registered and authorized students, approved companies, faculty members, and the central internship unit.

One such permissioned blockchain platform is Hyperledger Fabric, which supports controlled multi-organization workflows. It provides membership services, channels, endorsement policies, and chaincode for implementing application-specific business logic \cite{androulaki2018fabric}. Another study also describes Fabric as an open-source permissioned framework with configurable consensus, membership services, and modular design for enterprise-level applications \cite{li2020survey}. Furthermore, Fabric addresses the performance and scalability limitations of public blockchain platforms using an execute–order–validate architecture \cite{cachin2016architecture}. This allows parallel transaction execution and makes Fabric suitable for StajChain, which requires controlled agreement workflows among multiple pre-registered and authorized actors.

\subsubsection{Hyperledger Fabric Performance Studies}
After selecting Hyperledger Fabric as a suitable permissioned blockchain platform, its performance characteristics should also be considered. Several studies have evaluated the performance of Hyperledger Fabric under different workloads and configurations. Nasir et al. compared Fabric v0.6 and v1.0 using execution time, latency, throughput, and scalability metrics \cite{nasir2018performance}. Thakkar et al. showed that parameters such as block size, endorsement policy, channels, resource allocation, and state database choice can affect throughput and latency \cite{thakkar2018fabric}. Moreover, Sukhwani et al. proposed a stochastic performance model for Hyperledger Fabric using Stochastic Reward Nets to analyze throughput, utilization, queue length, and system bottlenecks \cite{sukhwani2018modeling}. 

Later, Guggenberger et al. demonstrated that Fabric can support many enterprise use cases, but configuration choices such as database type, endorsement policy, private data usage, transaction size, and network conditions highly affect throughput and latency \cite{guggenberger2022performance}. Furthermore, Abbasi et al. benchmarked different Fabric network topologies and reported that some topology changes improved throughput by more than 150\%, while latency and resource consumption increased under higher transaction loads \cite{abbasi2025benchmarking}. These studies show that Fabric performance depends on configuration and workload. This is acceptable for StajChain because this project focuses on controlled administrative workflows rather than high-frequency transaction processing. 

\subsection{Positioning of StajChain}

The reviewed studies show that blockchain has been used for document management, administrative workflows, and academic certificate verification. While document-based studies focus on secure record keeping, version history, and transparent workflow actions \cite{das2022management,azzam2023OCRblockchain,rahat2025eviblock,krogsboll2020smart}, education-based studies focus on issuing, storing, sharing, or verifying existing academic records such as certificates and learning credentials \cite{turkanovic2018eductx,han2018educationrecords,grather2018lifelong,dash2025hyperledger}. These studies motivated the StajChain project because they show that blockchain can support trust, traceability, and verification in institutional processes. StajChain has similar requirements, such as preserving the integrity of agreement records and making previous actions visible to authorized parties. Blockchain can support these requirements by recording important agreement actions in an immutable and transparent way. 

Although some internship-related blockchain studies exist, they do not directly address the internship agreement approval lifecycle. For example, Widayanti et al. used blockchain for internship grade conversion and data governance, and their focus is on reliable evaluation data \cite{widayanti2024internship}. On the other hand, StajChain addresses a different problem by focusing on the internship agreement process. It aims to solve the integrity, traceability, and manual verification problems in this process using a permissioned blockchain-based workflow.

The need for this type of solution can be seen in the current ITU internship process. In the current ITU internship process, students create internship records through the university portal, download the required internship agreement documents, obtain the necessary student and company signatures manually, and upload the approved documents back to the system for faculty control \cite{ituStajGenelEsaslar,ituKariyerStajMerkezi}. Moreover, ITU internship rules state that internship documents should not be modified after they are downloaded from the system. Also, incomplete or incorrect documents can cause the internship to be cancelled \cite{ituKariyerStajMerkezi}. These rules show that the internship process includes strict document and approval requirements. However, since the document signing and exchange steps between the student and the company are manual and mostly left to the student, document changes or unauthorized signatures are difficult to detect through manual control of the faculty internship committee. This can cause delays in the approval process. Moreover, it can result in administrative and financial problems for the university if an incorrect agreement is approved.

StajChain is designed to improve this process by adding blockchain-based integrity, traceability, and controlled execution. The system manages the full lifecycle of an internship agreement, including agreement creation, sequential approvals, rejection handling, activation, completion, and ledger history queries. Furthermore, previous actions on an agreement can be viewed more transparently, and companies can be reviewed and approved before they become participants in the system.

The choice of Hyperledger Fabric is also related to these requirements. A permissioned blockchain is more suitable for the StajChain project since it involves pre-registered and authorized participants such as students, companies, faculty members, and the central internship unit. Hyperledger Fabric is a suitable option for this permissioned structure because it provides membership services, endorsement policies, channels, and chaincode. It also supports controlled multi-party workflows and application-specific business rules through smart contracts \cite{androulaki2018fabric,li2020survey}. For the performance, previous studies show that Fabric performance depends on workload and configuration \cite{thakkar2018fabric,guggenberger2022performance,abbasi2025benchmarking}. This is acceptable for StajChain because this project mainly focuses on the internship agreement management process, where transparency, auditability, correctness, and controlled access are more important than high-frequency transaction processing.

Table~\ref{tab:feature_comparison} summarizes the feature and positioning differences between StajChain and the most related studies. Performance-related results will be mentioned in detail in Section \ref{comp_eval}.

\begin{table}[H]
\centering
\small
\caption{Feature-Based Comparison of StajChain with Related Studies}
\label{tab:feature_comparison}
\hspace*{-0.5cm}
\begin{tabular}{|p{3.1cm}|c|c|c|c|}
\hline
\textbf{System} &
\textbf{Integrity} &
\textbf{Access Control} &
\textbf{Agreement Lifecycle} &
\textbf{Context} \\ \hline

Das et al. \cite{das2022management} &
Yes & Yes & No & Document \\ \hline

Krogsbøll et al. \cite{krogsboll2020smart} &
Yes & Yes & No & Administrative \\ \hline

EduCTX \cite{turkanovic2018eductx} &
Yes & Partial & No & Education \\ \hline

Blockchain for Education \cite{grather2018lifelong} &
Yes & Yes & No & Education \\ \hline

Widayanti et al. \cite{widayanti2024internship} &
Yes & Partial & No & Internship \\ \hline

StajChain &
Yes & Yes & Yes & Internship \\ \hline

\end{tabular}
\end{table}

In the table, "Yes" means that the feature is directly addressed, "Partial" means that it is indirectly or partly addressed, and "No" means that it is not the main focus of the study.

\newpage

\section{Developed Approach and System Model}
\label{sys_model}

\subsection{Data Model}
\label{data_model}

The data model of the StajChain project is separated into three main parts. It includes the off-chain application database, the on-chain ledger data, and the Fabric identity storage. Each part has different responsibilities in the system. 

The off-chain application database stores data that is required by the web application, such as users, companies, company requests, and agreement metadata. Meanwhile, the on-chain ledger stores the core internship agreement records and the state changes. On the other hand, the Fabric identity storage keeps all the certificates and the private keys that are used when the backend submits transactions to the Hyperledger Fabric network.

The figure shows the general data model of the system:

\begin{figure}[H]
    \centering
    \includegraphics[width=1\linewidth]{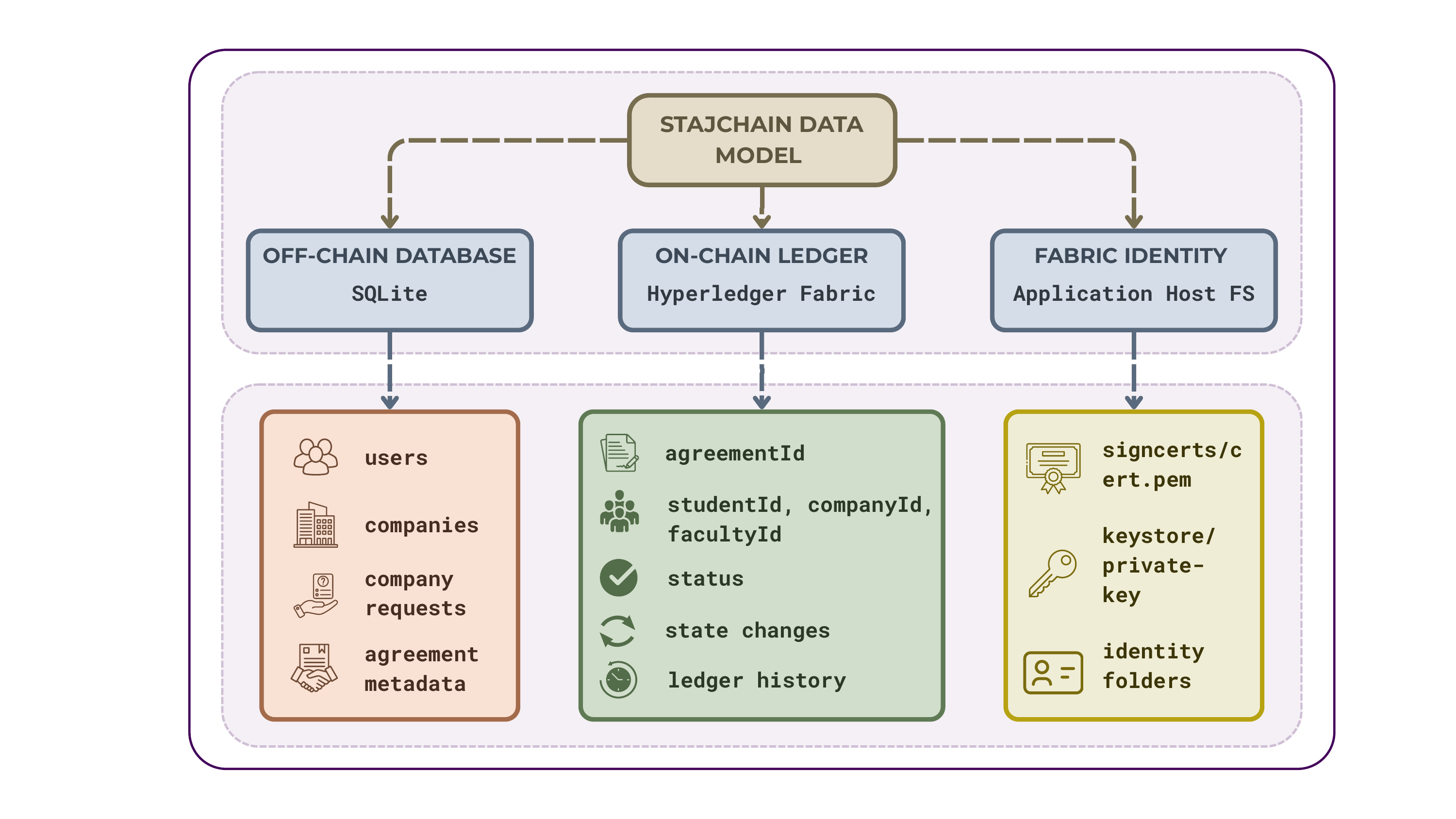}
    \caption{General data model of the StajChain system}
    \label{fig:datamodeloverview}
\end{figure}

The table below summarizes the data groups and their storage locations. In this table, SQLite represents the off-chain application database, while the Hyperledger Fabric ledger represents the on-chain blockchain storage. On the other hand, Fabric identities are stored separately in the application host file system.

\begin{table}[H]
\centering
\caption{Data storage summary of the StajChain system}
\label{tab:data_storage_summary}
\begin{tabularx}{\textwidth}{|p{3.2cm}|p{3.4cm}|X|}
\hline
\textbf{Data Group} & \textbf{Storage Location} & \textbf{Purpose} \\ \hline
Users and roles & SQLite database & Stores login data, role information, profile data, and Fabric identity references. \\ \hline
Companies & SQLite database & Stores approved company records used in agreement creation and company login. \\ \hline
Company requests & SQLite database & Stores company registration requests before approval by the central internship unit. \\ \hline
Agreement metadata & SQLite database & Stores agreement-related metadata and scheduling fields linked to the on-chain agreement ID. \\ \hline
Internship agreement records & Hyperledger Fabric ledger & Stores core agreement data, approval state, and lifecycle state transitions. \\ \hline
Agreement history & Hyperledger Fabric ledger & Stores previous versions and state changes of agreement records. \\ \hline
Fabric identities & Application host file system & Stores certificates and private keys used by the backend to submit Fabric transactions. \\ \hline
\end{tabularx}
\end{table}

\subsubsection{Off-Chain Application Database}
\label{appl_db}
The off-chain application layer stores user-related and company-related data in a relational database, SQLite. This layer is responsible for authentication, role-based authorization, and company management. 

The application layer consists of multiple related tables, including \texttt{users}, \texttt{companies}, \texttt{company\_requests}, and \texttt{agreement\_metadata}.

These tables are connected with application-level identifiers. The \texttt{users} table stores the login and role information of all actors. \texttt{companies} table stores all approved companies, while \texttt{company\_requests} table stores company requests before approval. The \texttt{agreement\_metadata} table is connected to the on-chain agreement record through the \texttt{agreement\_id} field.

\subsubsection*{Users Table}

The \texttt{users} table is the core data structure in the application layer and is responsible for handling authentication and role-based authorization. It stores login credentials, role information, and user profile attributes. Users include students, company users, faculty staff who are members of the internship committee, and the central internship unit. In the current prototype dataset, faculty-side actions are represented through faculty user accounts, and the provided demo configuration includes one faculty account.

In the current design, users do not register themselves through the system. User records, such as students, faculty users, and the central internship unit, are predefined during the configuration phase. Meanwhile, company user accounts are created later when a company registration request is approved. 

When a user performs a login, they are authenticated using their email address or username, together with their password, where the credentials are stored in the SQLite database. Passwords are stored as bcrypt hashes. After successful authentication, the backend generates a JWT token that includes the user's role, identity, and profile attributes. Then, \texttt{fabric\_identity} value is used to establish a connection to the Fabric layer.

Each user in the system is represented with the following attributes:

\begin{itemize}
    \item \texttt{name}, \texttt{surname}
    \item \texttt{username}: used for user login
    \item \texttt{email}: used for user login
    \item \texttt{password\_hash}: stores bcrypt hashed passwords
    \item \texttt{role}: represents the type of user (student, company, faculty, central unit)
    \item \texttt{entity\_id}: represents the unique identity of the user
    \item \texttt{fabric\_identity}: links the user to a blockchain identity
    \item \texttt{is\_active}: indicates whether the user account is active
    \item \texttt{created\_at}: stores when the user record is created
    \item \texttt{updated\_at}: stores when the user record is last updated
    \item \texttt{faculty\_id}, \texttt{faculty\_name}: store the faculty information of the student
    \item \texttt{department\_code}, \texttt{department\_name}: store the department information of the student
    \item \texttt{completed\_credits}: store the number of course credits completed by a student
\end{itemize}

Table~\ref{tab:users_table} below shows the logical structure of the users table: 

\begin{table}[H]
\centering
\caption{Logical Schema of the Users Table}
\label{tab:users_table}
\begin{tabularx}{\textwidth}{|l|l|X|}
\hline
\textbf{Column Name} & \textbf{Data Type} & \textbf{Description / Constraint} \\ \hline
id & INTEGER & Primary key, autoincrement \\ \hline
email & TEXT & Not null, unique \\ \hline
username & TEXT & Unique \\ \hline
password\_hash & TEXT & Not null \\ \hline
role & TEXT & Not null, user's role \\ \hline
entity\_id & TEXT & Not null, unique \\ \hline
fabric\_identity & TEXT & Not null, unique \\ \hline
is\_active & INTEGER & Not null, default value is 1 \\ \hline
name & TEXT & User name \\ \hline
surname & TEXT & User surname \\ \hline
faculty\_id & TEXT & Faculty identifier \\ \hline
faculty\_name & TEXT & Faculty name \\ \hline
department\_code & TEXT & Department code \\ \hline
department\_name & TEXT & Department name \\ \hline
completed\_credits & INTEGER & Number of completed credits \\ \hline
created\_at & TEXT & Not null, creation timestamp \\ \hline
updated\_at & TEXT & Not null, update timestamp \\ \hline
\end{tabularx}
\end{table}

The \texttt{username} field is optional, and uniqueness is enforced only for non-null values.

This design supports authentication at the backend layer and enables the system to enforce role-based access control within the internship workflow. In addition to the \texttt{role}, the \texttt{entity\_id} field is important to distinguish between specific actors. Moreover, the faculty-related, department-related, and completed credit information are used by the backend to enforce business rules such as minimum credit requirements, internship start-date constraints, department-specific internship constraints, and allowed internship fields.

In the current prototype, validation is shared between the backend and the smart contract. The backend validates user-related and application-level constraints before submitting a transaction. Meanwhile, the smart contract enforces role ownership, state transitions, and ledger-level agreement rules.   
Each user has a corresponding Fabric identity stored in the file system, referenced by the \texttt{fabric\_identity} field in the \texttt{users} table. There is a directory in the system that represents each Fabric identity.

Meanwhile, detailed company records are managed separately in the \texttt{companies} table. The company login credentials and Fabric identity references are still represented in the \texttt{users} table. However, in the current system, a company must be pre-registered to be part of an internship agreement. To manage this, companies are introduced into the system through a controlled request workflow. If students cannot find the company they want in the list, they can submit company information, which is stored in the \texttt{company\_requests} table.

\subsubsection*{Company Requests Table}

This table manages the lifecycle of company registration requests. Each request includes the submitted company information, the requesting student's entity identifier, the review status, an optional rejection reason, and the central unit identifier that reviewed the request. The central unit reviews these requests and either approves or rejects them. Using this method, controlled onboarding of companies into the system is enabled, and the risk of unverified or unauthorized company records can be reduced.

When a company request is approved, the backend generates a unique company identifier and username, creates a temporary password, provisions a Fabric identity for the company through the Fabric CA workflow, creates the company record, and creates the corresponding application-level user account. Thus, the company request approval performs database-side onboarding and provisions blockchain identity in a single controlled step.

Table~\ref{tab:company_requests_table} below shows the logical structure of the company requests table:

\begin{table}[H]
\centering
\caption{Logical Schema of the Company Requests Table}
\label{tab:company_requests_table}
\begin{tabularx}{\textwidth}{|l|l|X|}
\hline
\textbf{Column Name} & \textbf{Data Type} & \textbf{Description / Constraint} \\ \hline
id & INTEGER & Primary key, autoincrement \\ \hline
company\_name & TEXT & Not null, company name \\ \hline
company\_address & TEXT & Not null, company address \\ \hline
company\_phone\_number & TEXT & Not null, company phone number \\ \hline
company\_fax\_number & TEXT & Not null, company fax number \\ \hline
company\_email & TEXT & Not null, company email \\ \hline
is\_public\_institution & INTEGER & Not null, indicates whether the company is a public institution \\ \hline
company\_title & TEXT & Not null, company title \\ \hline
company\_iban & TEXT & Not null, company IBAN \\ \hline
company\_bank\_name & TEXT & Not null, bank name \\ \hline
company\_bank\_branch\_code & TEXT & Not null, bank branch code \\ \hline
company\_bank\_branch\_name & TEXT & Not null, bank branch name \\ \hline
company\_registration\_number & TEXT & Company registration number \\ \hline
company\_tax\_identification\_number & TEXT & Company tax identification number \\ \hline
requested\_by\_student\_id & TEXT & Not null, ID of the student who submitted the request \\ \hline
request\_status & TEXT & Not null, current status of the company request \\ \hline
rejection\_reason & TEXT & Reason for rejection, if the request is rejected \\ \hline
reviewed\_by\_central\_id & TEXT & ID of the central internship unit user who reviewed the request \\ \hline
created\_at & TEXT & Not null, creation timestamp \\ \hline
updated\_at & TEXT & Not null, update timestamp \\ \hline
\end{tabularx}
\end{table}

The request status can take values such as \texttt{PENDING}, \texttt{APPROVED}, or \texttt{REJECTED}.

When a request is approved, the system performs the following steps:

\begin{itemize}
    \item Creates a new company record in the \texttt{companies} table
    \item Creates a corresponding user account with role \texttt{company}
    \item Links a Fabric identity to the company
    \item Generates temporary login credentials
    \item Marks the request as \texttt{APPROVED}
\end{itemize}

At the end of the process, each approved company has both an application-level identity and a blockchain identity.

\subsubsection*{Companies Table}

All active approved company records are stored in \texttt{companies} table. This table contains company identifiers, usernames, contact information, public institution status, financial details, registration or tax information, active status, and the corresponding Fabric identity of the company.

Table~\ref{tab:companies_table} below shows the logical structure of the companies table:

\begin{table}[H]
\centering
\caption{Logical Schema of the Companies Table}
\label{tab:companies_table}
\begin{tabularx}{\textwidth}{|l|l|X|}
\hline
\textbf{Column Name} & \textbf{Data Type} & \textbf{Description / Constraint} \\ \hline
id & INTEGER & Primary key, autoincrement \\ \hline
company\_id & TEXT & Not null, unique company identifier \\ \hline
username & TEXT & Not null, unique company username \\ \hline
company\_name & TEXT & Not null, company name \\ \hline
company\_address & TEXT & Not null, company address \\ \hline
company\_phone\_number & TEXT & Not null, company phone number \\ \hline
company\_fax\_number & TEXT & Not null, company fax number \\ \hline
company\_email & TEXT & Not null, company email \\ \hline
is\_public\_institution & INTEGER & Not null, indicates whether the company is a public institution \\ \hline
company\_title & TEXT & Not null, company title \\ \hline
company\_iban & TEXT & Not null, company IBAN \\ \hline
company\_bank\_name & TEXT & Not null, bank name \\ \hline
company\_bank\_branch\_code & TEXT & Not null, bank branch code \\ \hline
company\_bank\_branch\_name & TEXT & Not null, bank branch name \\ \hline
company\_registration\_number & TEXT & Company registration number \\ \hline
company\_tax\_identification\_number & TEXT & Company tax identification number \\ \hline
is\_active & INTEGER & Not null, default value is 1 \\ \hline
fabric\_identity & TEXT & Fabric identity assigned to the company \\ \hline
created\_at & TEXT & Not null, creation timestamp \\ \hline
updated\_at & TEXT & Not null, update timestamp \\ \hline
\end{tabularx}
\end{table}

While the core internship agreement is stored on-chain, related agreement metadata is also stored in the \texttt{agreement\_metadata} table at the application layer.

\subsubsection*{Agreement Metadata Table}
The Agreement Metadata table stores off-chain agreement-related metadata about internship agreements. It keeps fields such as internship type, internship field, working days, weekly schedule, weekly working day count, and total working days. The table is linked to the on-chain agreement by the \texttt{agreement\_id} field. 

The Agreement Metadata table includes:

\begin{itemize}
    \item \texttt{agreement\_id}: uniquely identifies the internship agreement and links the metadata record to the on-chain agreement
    \item \texttt{internship\_type}: stores whether the internship is mandatory or voluntary
    \item \texttt{internship\_field}: stores the technical field of internship (ex. YAZILIM for the computer engineering department)
    \item \texttt{working\_days}: stores the fixed weekly working pattern used during the internship period
    \item \texttt{weekly\_schedule}: stores an optional flexible weekly schedule as serialized JSON, where each array represents the working days for a specific week
    \item \texttt{weekly\_working\_day\_count}: stores the number of selected working days per week for fixed schedules, or the maximum weekly working day count for flexible schedules
    \item \texttt{total\_working\_days}: stores the total number of actual working days
    \item \texttt{created\_at}: stores when the metadata record is created
    \item \texttt{updated\_at}: stores when the metadata record is last updated
\end{itemize}

The system supports two scheduling models. The \texttt{workingDays} field represents the fixed weekly schedule model, where the same selected days are repeated during the internship period. The \texttt{weeklySchedule} field represents the flexible weekly schedule model, where each week can have a different set of working days.

Table~\ref{tab:agreement_metadata_table} below shows the logical structure of the agreement metadata table:

\begin{table}[H]
\centering
\caption{Logical Schema of the Agreement Metadata Table}
\label{tab:agreement_metadata_table}
\begin{tabularx}{\textwidth}{|l|l|X|}
\hline
\textbf{Column Name} & \textbf{Data Type} & \textbf{Description / Constraint} \\ \hline
agreement\_id & TEXT & Primary key, on-chain agreement ID \\ \hline
internship\_type & TEXT & Not null, internship type \\ \hline
internship\_field & TEXT & Not null, internship field \\ \hline
working\_days & TEXT & Not null, fixed working days \\ \hline
weekly\_schedule & TEXT & Flexible schedule as JSON \\ \hline
weekly\_working\_day\_count & INTEGER & Not null, weekly day count \\ \hline
total\_working\_days & INTEGER & Not null, total working days \\ \hline
created\_at & TEXT & Not null, creation timestamp \\ \hline
updated\_at & TEXT & Not null, update timestamp \\ \hline
\end{tabularx}
\end{table}

\subsubsection{On-chain Ledger Layer}
\label{ledger_layer}

The on-chain ledger layer stores the critical internship agreement data. In Hyperledger Fabric, the ledger keeps committed transactions, and the world state keeps the latest version of each agreement record. In this project, each object is stored on the ledger as a structured object using its \texttt{agreementId} as the key. Smart contracts update the Internship Agreement Object when an agreement is approved, rejected, activated, or completed. Moreover, all status changes are recorded in the ledger history.

The core data structure of the blockchain layer is the Internship Agreement object. This object represents the agreement record that passes through the internship approval lifecycle. The \texttt{agreementId} attribute uniquely identifies each agreement. The object also includes the attributes below: 

\begin{itemize}
    \item \texttt{studentId}: Identifier of the student 
    \item \texttt{companyId}: Identifier of the company
    \item \texttt{facultyId}: Identifier of the faculty that the student is registered in
    \item \texttt{startDate}, \texttt{endDate}: Internship duration
    \item \texttt{internshipType}: Type of internship, such as mandatory or voluntary
    \item \texttt{internshipField}: Technical field of the internship
    \item \texttt{workingDays}: Selected working days for fixed weekly schedules
    \item \texttt{weeklySchedule}: Selected working days for flexible weekly schedules
    \item \texttt{weeklyWorkingDayCount}: Number of working days per week, or maximum weekly working day count for flexible schedules
    \item \texttt{totalWorkingDays}: Total number of actual working days
    \item \texttt{status}: Current state of the agreement
    \item \texttt{rejectedAt}: Timestamp that stores when the agreement is rejected
    \item \texttt{rejectedByRole}: Role of the actor who rejected the agreement
    \item \texttt{rejectionReason}: Optional reason if rejected
    \item \texttt{completedAt}: Timestamp that stores when the agreement is completed
    \item \texttt{createdAt}: Timestamp that stores when the agreement record is created
    \item \texttt{updatedAt}: Timestamp that stores when the agreement record is last updated
\end{itemize}

The \texttt{status} field is one of the main attributes of the agreement object because it controls the workflow. It can take \texttt{CREATED}, \texttt{STUDENT\_APPROVED}, \texttt{COMPANY\_APPROVED}, \texttt{FACULTY\_APPROVED}, \texttt{ACTIVE}, \texttt{COMPLETED}, or \texttt{REJECTED} values. These states are changed only by smart contract functions. The agreement status cannot be updated directly from the application database.

Smart contracts also validate the agreement data before storing it on-chain. The smart contract checks the date range, internship type, internship field, working day format, weekly schedule format, and total working day count during agreement creation. The total number of working days is recalculated by the smart contract and compared with the submitted value. This prevents inconsistent internship schedule data from being stored on the ledger. 

Agreement access and rejection rules depend on both the actor identity and the current agreement status. These rules are enforced by the smart contract and are explained in detail in Section \ref{str_model}.

In addition to the main agreement object, the smart contract creates composite-key index records for listing agreements by actor. These indices are auxiliary ledger records. They do not replace the main Internship Agreement object. Instead, they are used to retrieve agreements related to a specific student, company, or faculty more easily.

\begin{itemize}
    \item \texttt{student\textasciitilde agreement}: Index for agreements of a student
    \item \texttt{company\textasciitilde agreement}: Index for agreements of a company
    \item \texttt{faculty\textasciitilde agreement}: Index for agreements of a faculty
\end{itemize}

Table~\ref{tab:onchain_field_groups} summarizes the main field groups of the on-chain agreement object.

\begin{table}[H]
\centering
\caption{Main field groups of the on-chain agreement object}
\label{tab:onchain_field_groups}
\begin{tabularx}{\textwidth}{|p{4cm}|X|}
\hline
\textbf{Field Group} & \textbf{Purpose} \\ \hline
Identity fields & Store identifiers such as \texttt{agreementId}, \texttt{studentId}, \texttt{companyId}, and \texttt{facultyId}. \\ \hline
Internship fields & Store internship duration, internship type, internship field, working days, weekly schedule, and total working days. \\ \hline
Workflow fields & Store the current \texttt{status} of the agreement and determine the next allowed action. \\ \hline
Rejection fields & Store rejection timestamp, rejecting actor role, and optional rejection reason. \\ \hline
Timestamp fields & Store creation, update, rejection, and completion timestamps. \\ \hline
Index records & Store auxiliary composite-key records for listing agreements by student, company, or faculty. \\ \hline
\end{tabularx}
\end{table}

The logical structure of the on-chain Internship Agreement object is as follows:

\begin{verbatim}
{
  "agreementId": "agreement-...",
  "studentId": "student123",
  "companyId": "companyB",
  "facultyId": "facultyENG",
  "startDate": "2026-07-01",
  "endDate": "2026-08-01",
  "internshipType": "MANDATORY",
  "internshipField": "YAZILIM",
  "workingDays": ["MON", "TUE", "WED", "THU", "FRI"],
  "weeklySchedule": null,
  "weeklyWorkingDayCount": 5,
  "totalWorkingDays": 24,
  "status": "CREATED",
  "createdAt": "2026-04-26T10:00:00.000Z",
  "updatedAt": "2026-04-26T10:00:00.000Z",
  "rejectedAt": null,
  "rejectedByRole": null,
  "rejectionReason": null,
  "completedAt": null
}
\end{verbatim}

\subsubsection{Fabric Identity Storage}
\label{ident_str}

Fabric identity storage is separated from both the SQLite database and the on-chain ledger. In this project, a Fabric identity means the certificate and private key files that represent a user in the Hyperledger Fabric network. These files are not stored in the SQLite application database or on the ledger. The SQLite database stores only the \texttt{fabric\_identity} reference in the \texttt{users} table. This identity is a reference to the certificate and key files stored in the application host file system. This reference points to a directory under the identity storage folder. These directories contain the user certificates and private keys.

When a user performs an action that requires a blockchain transaction, the backend uses the \texttt{fabric\_identity} value to locate the corresponding identity folder. Then, the certificate is loaded from the \texttt{signcerts} directory and the private key is loaded from the \texttt{keystore} directory to establish a connection to the blockchain network. The certificate is used as the X.509 identity of the user, while the private key is used to sign transaction proposals.

The general structure of a Fabric identity directory is as follows:

\begin{verbatim}
stajchain-identities/
  student123/
    signcerts/
      cert.pem
    keystore/
      <private-key-file>
    cacerts/
    user/
\end{verbatim}

Users do not directly interact with these Fabric credentials. The backend performs blockchain operations for the authenticated user by using the Fabric identity referenced in the database. At the smart contract level, authorization depends on the user attributes such as \texttt{role} and \texttt{id}, which are embedded in the Fabric certificate. The \texttt{role} attribute determines the type of actor, such as student, company, faculty, or central unit. The \texttt{id} attribute is used to compare the caller with agreement fields such as \texttt{studentId}, \texttt{companyId}, or \texttt{facultyId}.

In the current implementation, Fabric identities are created during the configuration phase for predefined users such as students, faculty staff, and the central internship unit. Company identities are created only when a company registration request is approved. After the identity is created, the directory name is stored as the \texttt{fabric\_identity} value of the corresponding user. The company registration process will be mentioned in detail in Section \ref{comp_reg}.

\subsection{Structural Model}
\label{str_model}

The StajChain system is organized around the frontend, backend, off-chain database, and Hyperledger Fabric blockchain network. Users interact with the system through the React frontend. Then, the frontend sends requests to the Node.js backend, which handles authentication, role-based access control, business rule validation, and blockchain gateway operations. The backend stores application-level data in the SQLite off-chain database and submits transactions to the Hyperledger Fabric network. Thus, the backend acts as a bridge between the off-chain database and the on-chain ledger.

\begin{figure}[H]
    \centering
    \includegraphics[width=1\linewidth]{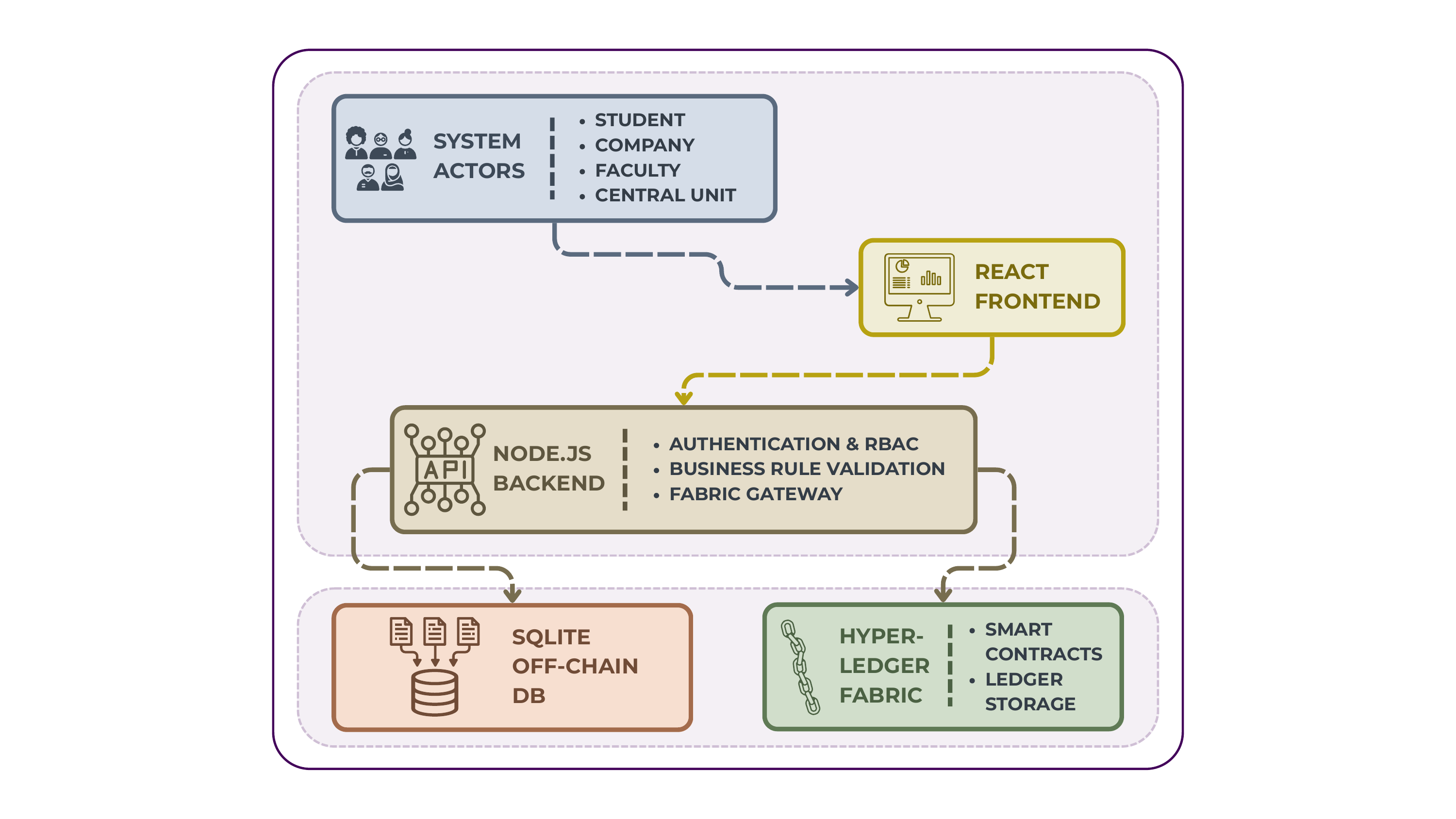}
    \caption{System architecture of the StajChain prototype}
    \label{fig:systemarchitecture}
\end{figure}

\subsubsection{Business Roles and Access Control}

The system defines four main actors, which are the student, the company, the faculty, and the central internship unit. These roles determine which actions a user can perform in the system. The backend first checks the authenticated user's role before allowing access to an API endpoint. Then, the smart contracts check the caller role, actor identity, and the current agreement status before changing anything on the ledger.

In the current implementation, a student can access only the agreements associated with that student, and a company can access only the agreements associated with that company. Moreover, a faculty user can access only the agreements related to that faculty. The central internship unit can access all agreements, but only when they have reached the \texttt{FACULTY\_APPROVED}, \texttt{ACTIVE}, or \texttt{COMPLETED} stages. Thus, agreement access is determined by both the identity of the actor and the current status of the agreement.

Furthermore, smart contracts enforce the approval order. Only the student can create an agreement and approve it initially. The company can approve the agreement only after student approval. Then, the faculty can approve it after the company approval. After faculty approval, the central internship unit can activate the agreement. At the end, the central unit marks the internship agreement as completed.

In a rejection scenario, a student can reject the agreement only if it is in \texttt{CREATED}, \texttt{STUDENT\_APPROVED}, or \texttt{COMPANY\_APPROVED} stages. The company can reject it only when the agreement is in the \texttt{STUDENT\_APPROVED} stage. The faculty can reject it only when the agreement is in the \texttt{COMPANY\_APPROVED} state. There is no rejection option for the central unit. It can only activate the agreement after faculty approval and complete it after the internship process ends. 

\subsubsection{Business Rules Enforced in the Prototype}

In addition to role-based access control, the prototype enforces several business rules during internship agreement creation and workflow execution. Some of these rules are checked by the backend before a transaction is submitted. Meanwhile, smart contracts check ledger-level rules. The enforced rules in the StajChain project are as follows:

\begin{itemize}
    \item Only students can create internship agreements.
    
    \item To create an agreement, the selected company must already exist in the system with a valid blockchain identity. Otherwise, the student has to submit a company registration request.
    
    \item A student must have completed at least 30 credits to create an internship agreement.
    
    \item The internship start date must be at least 15 days later than the current date. 
    
    \item The internship start date cannot be later than the end date.
    
    \item The internship duration must satisfy the minimum duration constraint according to the department rules.
    
    \item The internship field must be compatible with the student's department. In the current prototype, only predefined internship fields are allowed for each supported department.
    
    \item The system prevents overlapping internship periods for the same student. Thus, a student cannot create a new internship agreement if its period overlaps with another valid internship record.
    
    \item A student can work between 3 and 6 days a week. If a weekly schedule is used that allows students to select different working days across weeks, each week in the schedule must also contain between 3 and 6 working days.
    
    \item Mandatory and voluntary internships have different constraints. For mandatory internships, the total working days must satisfy the minimum requirement, which is specific to each department. Also, the student cannot exceed the department-specific maximum number of mandatory internships.
    
    \item A student cannot repeat a mandatory internship in the same internship field.
    
    \item For voluntary internships, the total working days must satisfy the minimum voluntary internship day requirement, and the student cannot exceed the total accumulated voluntary internship day limit.
    
    \item Approval order is pre-defined and cannot be changed. After an agreement is created, only the student can perform the first approval. Then, only the related company can approve after the student's approval. Later, only the related faculty can approve. After faculty approval, only the central internship unit can activate the agreement and can mark an active agreement as completed.

    \item Agreement visibility is restricted by role and agreement state. Students, companies, and faculty users can access only their own related agreements. The central internship unit can access agreements only at predefined stages.
    
    \item Rejections are controlled by role and agreement state. A student, company, or faculty can reject an agreement only at the permitted stages for that role. 
\end{itemize}

\subsubsection{User Authentication}
\label{user_auth}
The workflow starts with user authentication. As mentioned in Section \ref{appl_db}, users do not register themselves through the system. Instead, users except companies are registered during the configuration phase. Companies are registered after the central internship unit approves a company registration request. During a user login, the backend checks the submitted credentials against the stored password hash. If authentication is successful, the backend creates a JWT token. It contains user's role, entity identifier, profile attributes, and Fabric identity reference.

\begin{tcolorbox}[
    breakable,
    colback=white,
    colframe=black,
    boxrule=0.5pt,
    arc=0pt,
    left=6pt,
    right=6pt,
    top=6pt,
    bottom=6pt
]

\begin{center}
\textbf{User Authentication}
\end{center}

\begin{algorithmic}[1]
\Require Login identifier, password
\Ensure JWT token or authentication error
\State Normalize the login identifier
\State Find the user by email or username in the SQLite database
\If{user does not exist}
    \State Reject the login request
\EndIf
\If{user account is inactive}
    \State Reject the login request
\EndIf
\State Compare the submitted password with the stored bcrypt hash
\If{password does not match}
    \State Reject the login request
\EndIf
\State Check whether the assigned Fabric identity directory exists
\If{Fabric identity does not exist}
    \State Reject the login request
\EndIf
\State Generate a JWT token with user role, entity identifier, profile attributes, and Fabric identity
\State Return the JWT token and user information
\end{algorithmic}

\end{tcolorbox}

\subsubsection{Fabric Transaction Invocation}
\label{fbr_invoke}
After authentication, the backend uses the authenticated user's Fabric identity to perform blockchain operations. The backend locates the identity directory and loads the certificate and private key. Later, the user can perform transactions on the Fabric ledger.

\begin{tcolorbox}[
    breakable,
    colback=white,
    colframe=black,
    boxrule=0.5pt,
    arc=0pt,
    left=6pt,
    right=6pt,
    top=6pt,
    bottom=6pt
]

\begin{center}
\textbf{Fabric Transaction Invocation}
\end{center}

\begin{algorithmic}[1]
\Require Authenticated user, transaction name, transaction parameters
\Ensure Transaction result or error
\State Read the \texttt{fabric\_identity} value from the authenticated user
\State Locate the corresponding identity directory in the file system
\State Load the certificate from the \texttt{signcerts} directory
\State Load the private key from the \texttt{keystore} directory
\State Create the Fabric Gateway identity and signer
\State Connect to the Fabric network and access the chaincode
\State Submit or evaluate the requested transaction
\State Close the gateway connection
\State Return the transaction result
\end{algorithmic}

\end{tcolorbox}

\subsubsection{Internship Agreement Creation}
\label{crt_agr}
One of the core processes of the system is the creation of internship agreements. Only a student can create an internship agreement. Before writing the agreement to the ledger, validation is divided between the backend and the smart contract. The backend validates the authenticated caller, company availability, student metadata, department-specific constraints, and application-level date rules. The smart contract validates ledger-side business rules such as date-range consistency, internship type and field consistency, working-day or weekly-schedule structure, total working day count, overlap prevention, and agreement ownership.

\begin{tcolorbox}[
    breakable,
    colback=white,
    colframe=black,
    boxrule=0.5pt,
    arc=0pt,
    left=6pt,
    right=6pt,
    top=6pt,
    bottom=6pt
]

\begin{center}
\textbf{Create Internship Agreement}
\end{center}

\begin{algorithmic}[1]
\Require Agreement information, student Fabric identity
\Ensure Created agreement or validation error
\State Backend checks whether the selected company exists in the approved company list
\If{company does not exist or does not have a Fabric identity}
    \State Stop agreement creation and require a company registration request
\EndIf
\State Backend submits the creation transaction to the smart contract
\State Read caller attributes from the Fabric certificate
\If{caller role is not student}
    \State Reject the transaction
\EndIf
\If{caller id is not equal to the submitted student id}
    \State Reject the transaction
\EndIf
\State Validate start date and end date
\State Normalize internship type and internship field
\If{weekly schedule is provided}
    \State Validate weekly schedule format
    \State Calculate total working days from weekly schedule
\Else
    \State Validate fixed working days
    \State Calculate total working days from date range and working days
\EndIf
\If{submitted total working days is not equal to calculated total}
    \State Reject the transaction
\EndIf
\State Check overlapping internship agreements of the student
\State Check mandatory and voluntary internship business rules
\State Create the internship agreement object with status \texttt{CREATED}
\State Store the agreement on-chain
\State Create composite-key index records for student, company, and faculty
\State Emit agreement creation event
\State Return the created agreement identifier and status
\end{algorithmic}

\end{tcolorbox}

\subsubsection{Company Registration}
\label{comp_reg}

During the internship creation phase, when a company is not already available in the system, students can submit a company request. The central internship unit reviews the request. If the request is approved, the backend generates a unique company identifier and username and creates a temporary password. Then, the backend provisions a Fabric identity for the company through the Fabric CA, creates the company record, and creates the application-level user account.

\begin{tcolorbox}[
    breakable,
    colback=white,
    colframe=black,
    boxrule=0.5pt,
    arc=0pt,
    left=6pt,
    right=6pt,
    top=6pt,
    bottom=6pt
]

\begin{center}
\textbf{Approve Company Registration Request}
\end{center}

\begin{algorithmic}[1]
\Require Company request id, central unit user
\Ensure Approved company and company credentials or error
\State Check whether the caller role is central unit
\State Read the company request from the database
\If{request does not exist or is not pending}
    \State Reject the operation
\EndIf
\State Review the company information in the request
\If{company information is not valid or not suitable}
    \State Reject the company request with a rejection reason
    \State Stop the approval process
\EndIf
\State Generate a unique company identifier
\State Generate a unique company username
\State Generate a temporary password
\State Hash the temporary password using bcrypt
\State Provision a Fabric identity for the company
\If{Fabric identity provisioning fails}
    \State Stop the approval process
\EndIf
\State Create a company record in the \texttt{companies} table
\State Create a company user account in the \texttt{users} table
\State Store the generated Fabric identity reference
\State Mark the company request as \texttt{APPROVED}
\State Return the company information and temporary credentials
\end{algorithmic}

\end{tcolorbox}

\subsubsection{Agreement Approval Workflow}
\label{agr_workflow}
After an agreement is created, it passes through student, company, and faculty approval stages. Each approval function is implemented separately in the smart contract. The smart contract checks the role and entity identifier of the caller before changing the agreement status.

\begin{tcolorbox}[
    breakable,
    colback=white,
    colframe=black,
    boxrule=0.5pt,
    arc=0pt,
    left=6pt,
    right=6pt,
    top=6pt,
    bottom=6pt
]

\begin{center}
\textbf{Agreement Approval Workflow}
\end{center}

\begin{algorithmic}[1]
\Require Agreement id, caller Fabric identity
\Ensure Updated agreement status or authorization error
\State Read caller role and id from the Fabric certificate
\State Read the agreement from the ledger
\If{caller is student}
    \State Check whether caller id matches \texttt{studentId}
    \If{agreement status is \texttt{CREATED}}
        \State Set status to \texttt{STUDENT\_APPROVED}
    \Else
        \State Reject the transaction
    \EndIf
\ElsIf{caller is company}
    \State Check whether caller id matches \texttt{companyId}
    \If{agreement status is \texttt{STUDENT\_APPROVED}}
        \State Set status to \texttt{COMPANY\_APPROVED}
    \Else
        \State Reject the transaction
    \EndIf
\ElsIf{caller is faculty}
    \State Check whether caller id matches \texttt{facultyId}
    \If{agreement status is \texttt{COMPANY\_APPROVED}}
        \State Set status to \texttt{FACULTY\_APPROVED}
    \Else
        \State Reject the transaction
    \EndIf
\Else
    \State Reject the transaction
\EndIf
\State Update the \texttt{updatedAt} timestamp
\State Store the updated agreement on-chain
\State Emit the corresponding agreement event
\State Return the updated agreement status
\end{algorithmic}

\end{tcolorbox}

After faculty approval, the central internship unit activates the agreement. This process is different from the approval functions. It represents that the internship agreement has passed the required approval stages, and the formal procedures can start. Those procedures include the insurance activation in a real-life scenario.

\begin{tcolorbox}[
    breakable,
    colback=white,
    colframe=black,
    boxrule=0.5pt,
    arc=0pt,
    left=6pt,
    right=6pt,
    top=6pt,
    bottom=6pt
]

\begin{center}
\textbf{Activate Internship Agreement}
\end{center}

\begin{algorithmic}[1]
\Require Agreement id, central unit Fabric identity
\Ensure Activated agreement or authorization error
\State Read caller role and id from the Fabric certificate
\If{caller role is not central}
    \State Reject the transaction
\EndIf
\If{caller id is not equal to the central unit identifier}
    \State Reject the transaction
\EndIf
\State Read the agreement from the ledger
\If{agreement status is not \texttt{FACULTY\_APPROVED}}
    \State Reject the transaction
\EndIf
\State Set status to \texttt{ACTIVE}
\State Update the \texttt{updatedAt} timestamp
\State Store the updated agreement on-chain
\State Emit agreement activation event
\State Return the activated agreement status
\end{algorithmic}

\end{tcolorbox}

The central internship unit marks the agreement as completed when the internship process is finished. The agreement can be completed only if it is already active. In a real-life scenario, this represents the insurance deactivation and can be done only when the internship finishes. However, due to time and demonstration constraints, in this implementation, the completion action can be performed directly after the agreement is activated.

\begin{tcolorbox}[
    breakable,
    colback=white,
    colframe=black,
    boxrule=0.5pt,
    arc=0pt,
    left=6pt,
    right=6pt,
    top=6pt,
    bottom=6pt
]

\begin{center}
\textbf{Complete Internship Agreement}
\end{center}

\begin{algorithmic}[1]
\Require Agreement id, central unit Fabric identity
\Ensure Completed agreement or authorization error
\State Read caller role and id from the Fabric certificate
\If{caller role is not central}
    \State Reject the transaction
\EndIf
\If{caller id is not equal to the central unit identifier}
    \State Reject the transaction
\EndIf
\State Read the agreement from the ledger
\If{agreement status is not \texttt{ACTIVE}}
    \State Reject the transaction
\EndIf
\State Set status to \texttt{COMPLETED}
\State Store the completion timestamp in \texttt{completedAt}
\State Update the \texttt{updatedAt} timestamp
\State Store the updated agreement on-chain
\State Emit agreement completion event
\State Return the completed agreement status
\end{algorithmic}

\end{tcolorbox}

\subsubsection{Internship Agreement Rejection}
\label{reject_agr}
The system also supports rejection operations. Rejection is controlled by the current status of the agreement and the role of the caller. As mentioned in Section \ref{ledger_layer}, a student can reject the agreement only when it is in \texttt{CREATED}, \texttt{STUDENT\_APPROVED}, or \texttt{COMPANY\_APPROVED} stages. The company can reject the internship agreement only when it is in \texttt{STUDENT\_APPROVED} stage, and faculty can reject it only when it is in \texttt{COMPANY\_APPROVED} state. When an agreement is rejected, the smart contract stores the rejection timestamp, rejecting role, and rejection reason.

\begin{tcolorbox}[
    breakable,
    colback=white,
    colframe=black,
    boxrule=0.5pt,
    arc=0pt,
    left=6pt,
    right=6pt,
    top=6pt,
    bottom=6pt
]

\begin{center}
\textbf{Reject Internship Agreement}
\end{center}

\begin{algorithmic}[1]
\Require Agreement id, rejection reason, caller Fabric identity
\Ensure Rejected agreement or authorization error
\State Read caller role and id from the Fabric certificate
\State Read the agreement from the ledger
\State Validate the rejection reason
\If{caller is student}
    \State Check whether caller id matches \texttt{studentId}
    \State Allow rejection only in \texttt{CREATED}, \texttt{STUDENT\_APPROVED}, or \texttt{COMPANY\_APPROVED} status
\ElsIf{caller is company}
    \State Check whether caller id matches \texttt{companyId}
    \State Allow rejection only in \texttt{STUDENT\_APPROVED} status
\ElsIf{caller is faculty}
    \State Check whether caller id matches \texttt{facultyId}
    \State Allow rejection only in \texttt{COMPANY\_APPROVED} status
\Else
    \State Reject the transaction
\EndIf
\State Set status to \texttt{REJECTED}
\State Store \texttt{rejectedAt}, \texttt{rejectedByRole}, and \texttt{rejectionReason}
\State Update the \texttt{updatedAt} timestamp
\State Store the updated agreement on-chain
\State Emit the corresponding rejection event
\State Return the rejected agreement status
\end{algorithmic}

\end{tcolorbox}

\subsection{Dynamic Model}
\label{dyn_model}

The dynamic model explains the behavior of the StajChain system during the internship agreement process by using workflow models, state machine diagrams, and sequence diagrams.

\subsubsection{Agreement Lifecycle}

The agreement lifecycle model presents the main approval flow of the internship agreement process. It shows the main actions from agreement creation to completion, including approval, rejection, and activation steps.

The figure below illustrates the agreement lifecycle model of the system:

\begin{figure}[H]
    \centering
    \includegraphics[width=1\linewidth]{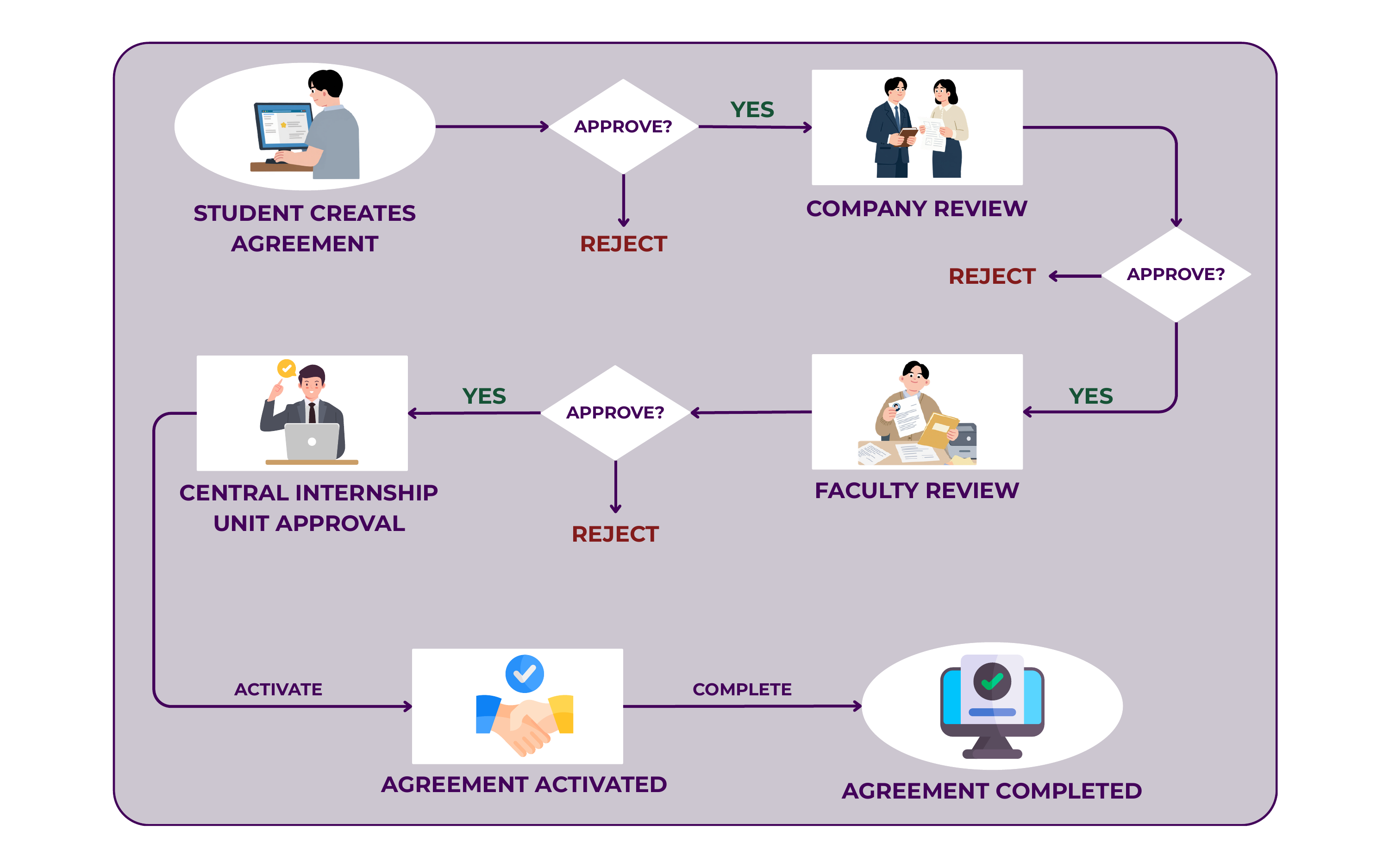}
    \caption{Agreement lifecycle model}
    \label{fig:agreementflow}
\end{figure}

\subsubsection{Agreement State Machine Diagram}

The main dynamic object in the system is the internship agreement. Its status changes according to the approvals or rejections performed by authorized actors. The system follows a sequential approval workflow including the student, company, faculty, and central internship unit.

The state machine diagram below focuses on the possible states of an agreement and the valid transitions between these states:

\begin{figure}[H]
    \centering
    \includegraphics[width=1\linewidth]{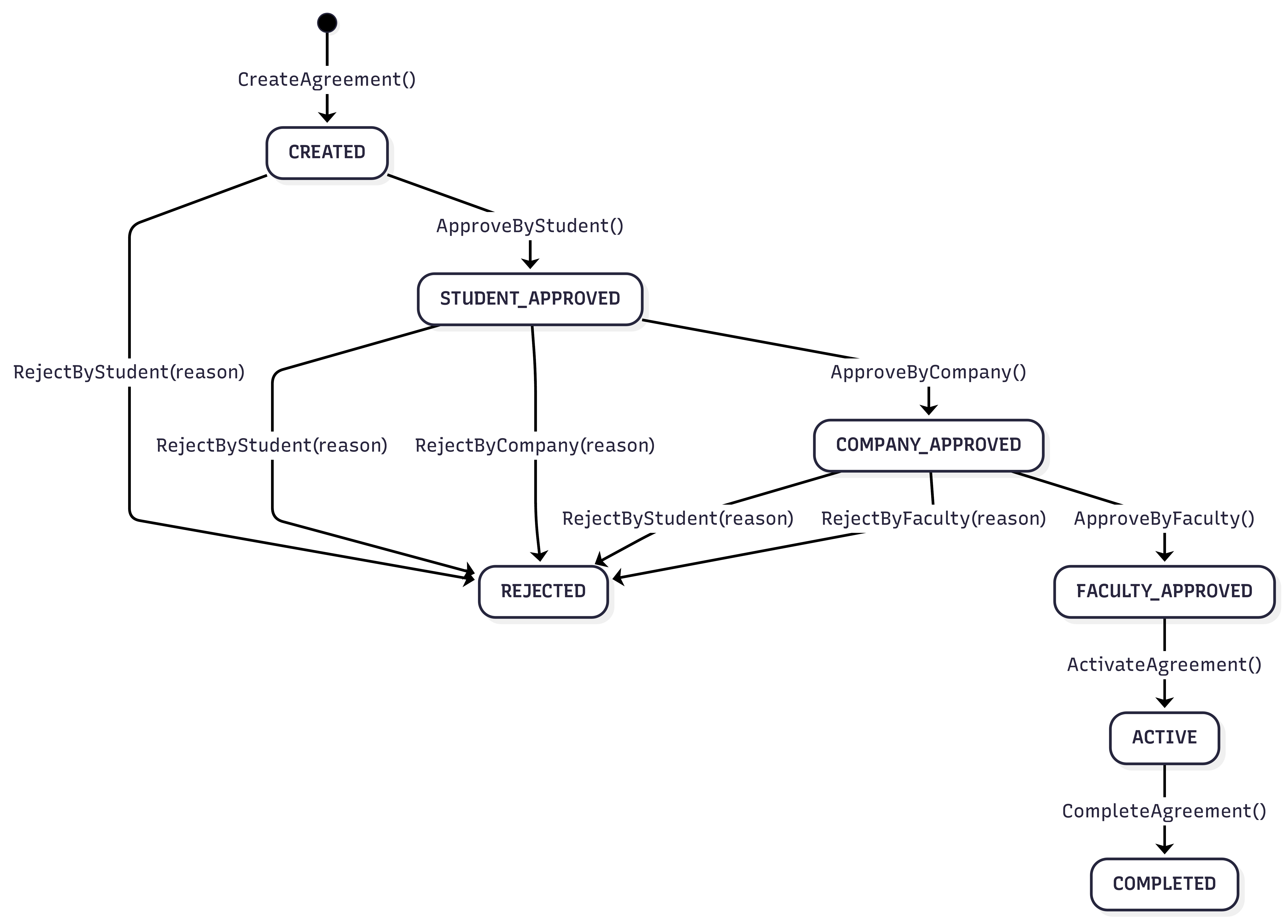}
    \caption{State Machine Diagram of the System}
    \label{fig:statemachinediag}
\end{figure}

The agreement lifecycle starts with the creation of the internship agreement by the student. In this state, the status of the agreement is \texttt{CREATED}. Then, the student approves or rejects this agreement. If the student approves, the status will be \texttt{STUDENT\_APPROVED}. After that, the company reviews the agreement and either approves or rejects it. If the company approves, the status will be \texttt{COMPANY\_APPROVED}. Later, the faculty either approves or rejects the agreement. After company approval, the status will be \texttt{FACULTY\_APPROVED}, and the central internship unit can activate the internship. Finally, an active internship can be completed by the central unit. 

At several stages, authorized actors may reject the agreement, which moves it to the REJECTED terminal state. The student can reject the agreement at \texttt{CREATED}, \texttt{STUDENT\_APPROVED}, or \texttt{COMPANY\_APPROVED} stages. The company can reject it only when it is in \texttt{STUDENT\_APPROVED} stage, and faculty can reject it only when it is in \texttt{COMPANY\_APPROVED} state. 

\newpage

\subsubsection{Overall Sequence Diagram}

\begin{figure}[H]
    \centering
    \hspace*{-2cm}
    \includegraphics[
        width=1.2\textwidth,
        height=0.9\textheight,
        keepaspectratio
    ]{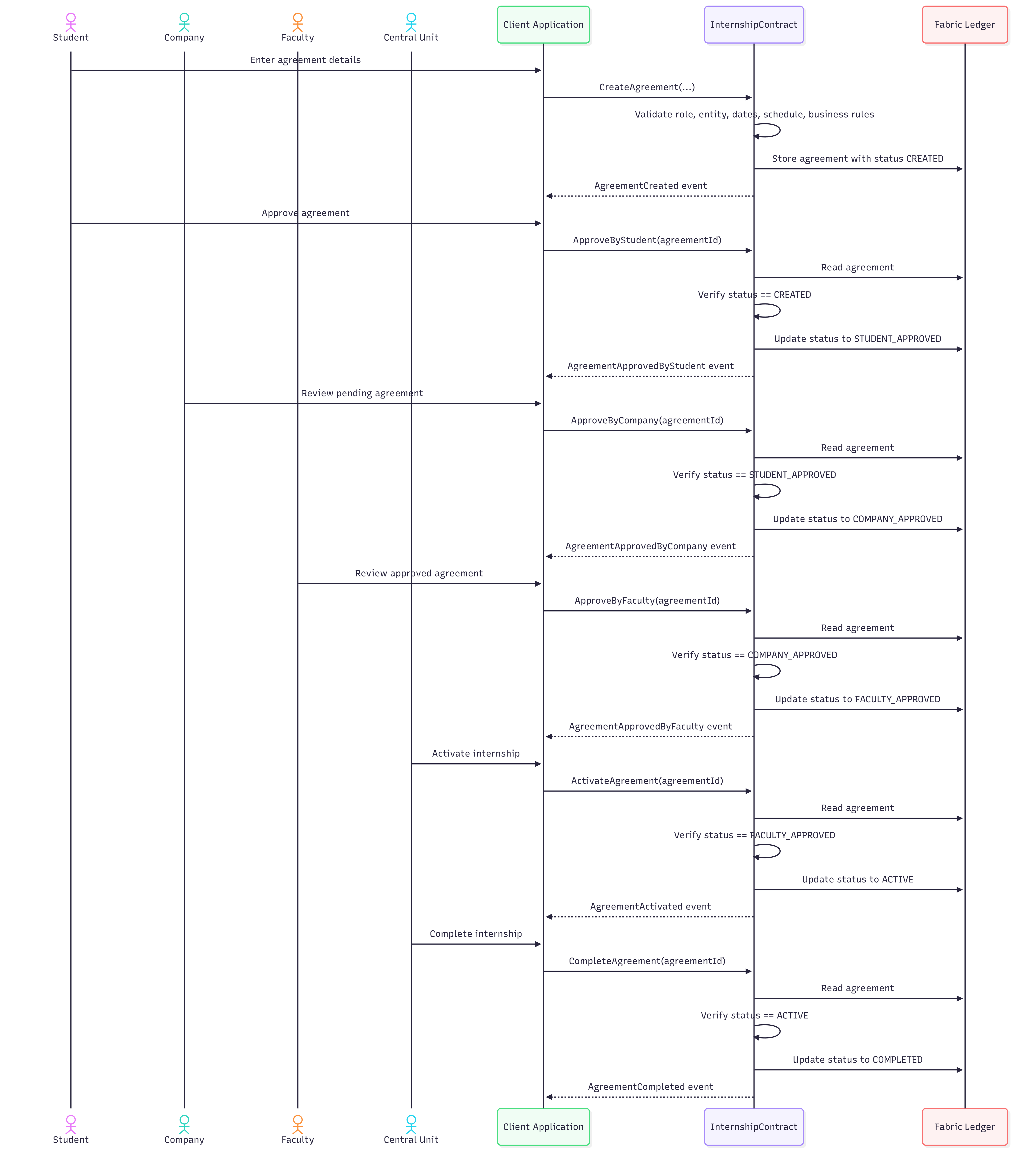}
    \caption{Overall Sequence Diagram of the System}
    \label{fig:sequencediag}
\end{figure}

Figure~\ref{fig:sequencediag} shows the overall sequence diagram of the system. This diagram can be difficult to read since it includes several actors and many interactions. Thus, the workflow is divided into separate component-based sequence diagrams for the student, company, faculty, and central unit. These diagrams explain each role's actions clearly.

\subsubsection{Student Sequence Diagram}
\begin{figure}[H]
    \centering
    \includegraphics[width=1\linewidth]{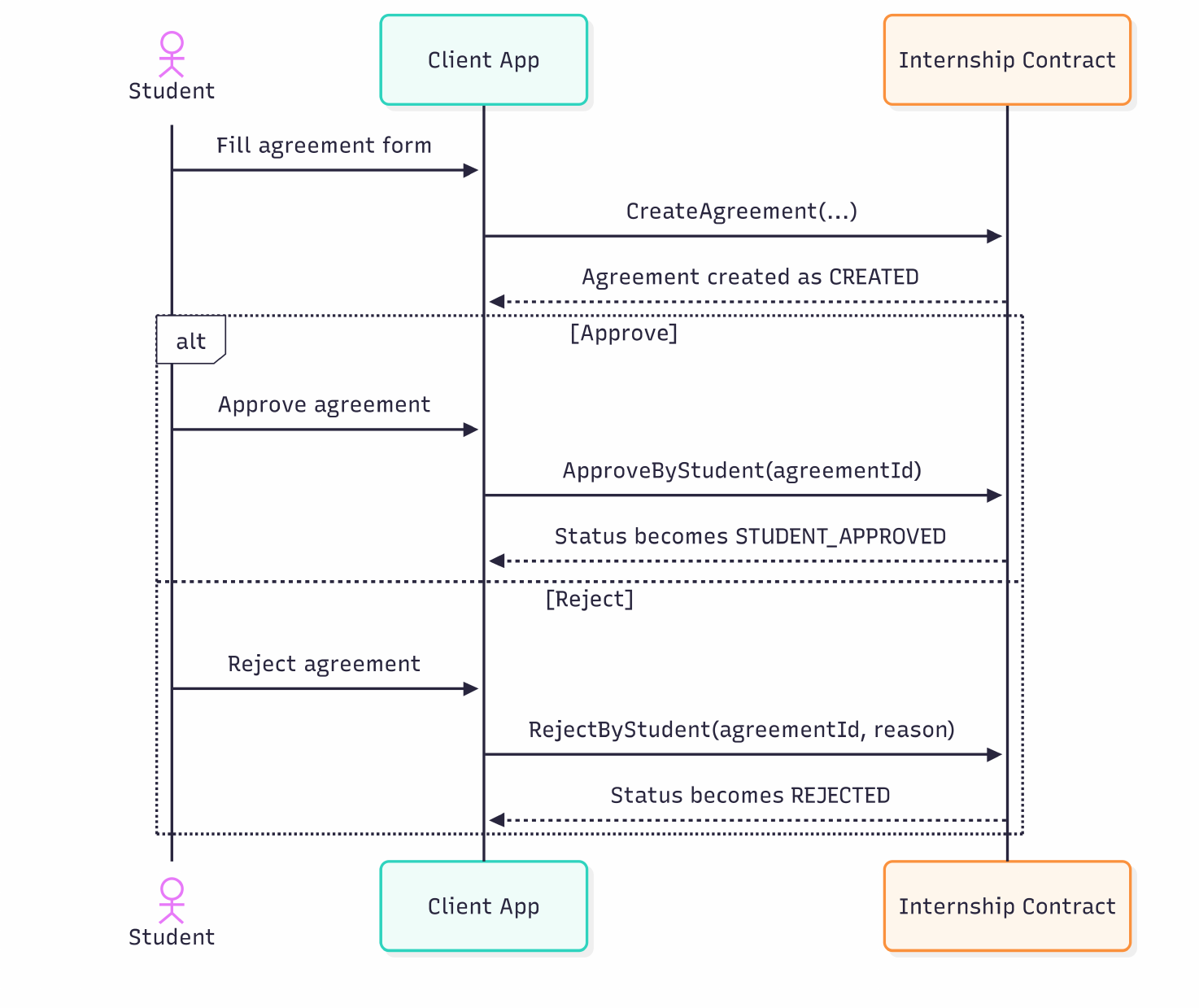}
    \caption{Student-Based Sequence Diagram of the System}
    \label{fig:studsequencediag}
\end{figure}

This sequence diagram is a component of the general sequence diagram in Figure \ref{fig:sequencediag}. This diagram illustrates the internship agreement workflow from the student’s perspective. The student initiates the internship agreement process by filling out the agreement form and creating an internship agreement. After creating the agreement, the student either approves or rejects the agreement. The rejection ensures that the student has full control over the initial submission and confirmation of the internship details. After the approval stage, the agreement becomes visible to the company for further evaluation. 

\subsubsection{Company Sequence Diagram}
\begin{figure}[H]
    \centering
    \includegraphics[width=1\linewidth]{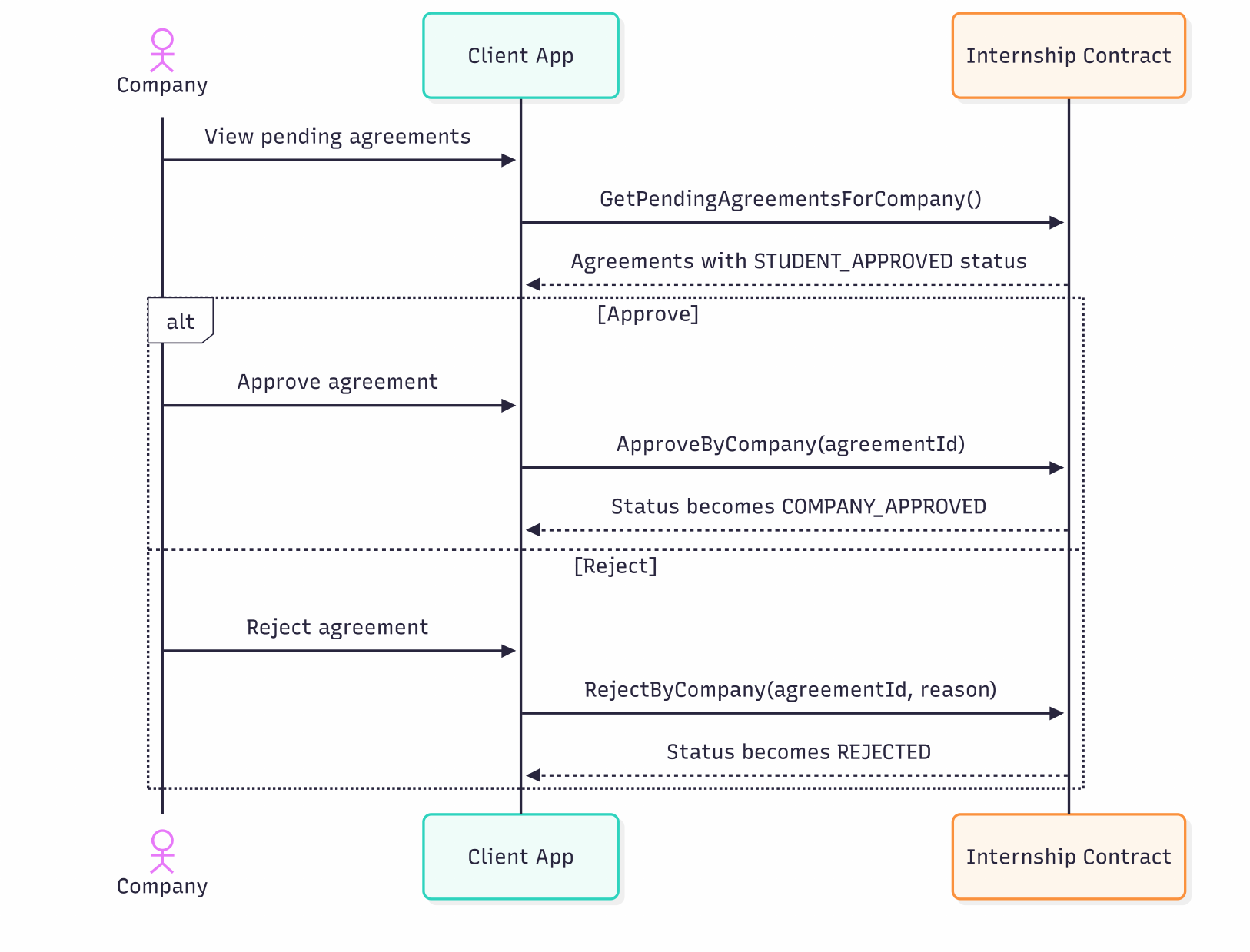}
    \caption{Company-Based Sequence Diagram of the System}
    \label{fig:compsequencediag}
\end{figure}

This sequence diagram is a component of the general sequence diagram in Figure \ref{fig:sequencediag}. This diagram illustrates the internship agreement workflow from the company’s perspective. A company can view all pending agreements that are related to that company and have the \texttt{STUDENT\_APPROVED} status. After review of the agreement details, the company has two options: approval or rejection. If the agreement is approved by the company, the agreement proceeds to faculty review. In contrast, if the company rejects the agreement, the process is terminated.

\subsubsection{Faculty Sequence Diagram}
\begin{figure}[H]
    \centering
    \includegraphics[width=1\linewidth]{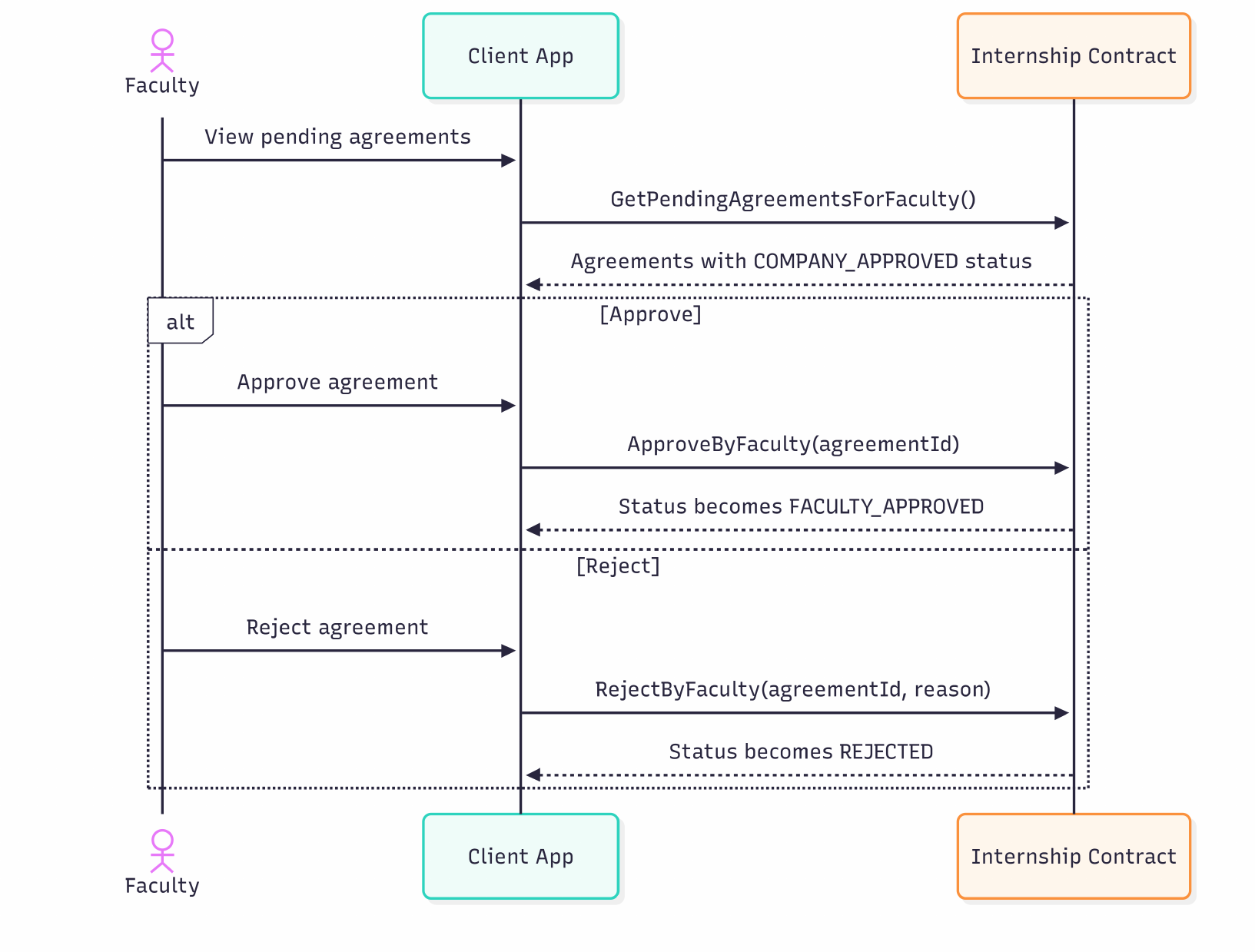}
    \caption{Faculty-Based Sequence Diagram of the System}
    \label{fig:facsequencediag}
\end{figure}

This sequence diagram is a component of the general sequence diagram in Figure \ref{fig:sequencediag}. This diagram illustrates the internship agreement workflow from the faculty’s perspective. A faculty can view all pending agreements that are related to that faculty and have the \texttt{COMPANY\_APPROVED} status. The faculty can either approve or reject the agreement after review. Approval makes the agreement ready for final activation by the central unit. If rejected, the agreement process terminates. This step ensures that all internships comply with academic regulations and institutional requirements.

\subsubsection{Central Unit Sequence Diagram}
\begin{figure}[H]
    \centering
    \includegraphics[width=1\linewidth]{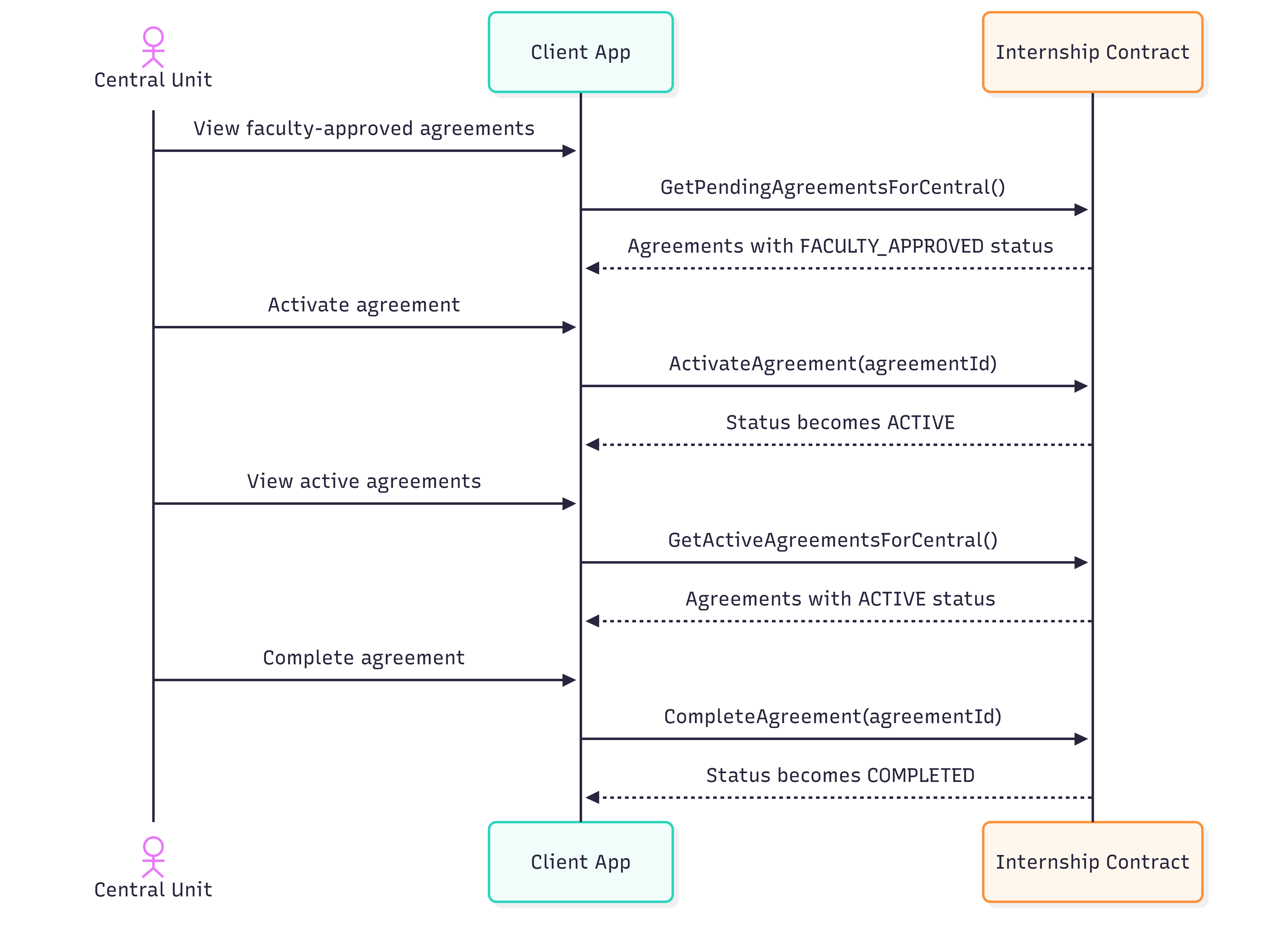}
    \caption{Central Unit-Based Sequence Diagram of the System}
    \label{fig:centralsequencediag}
\end{figure}

This sequence diagram is a component of the general sequence diagram in Figure \ref{fig:sequencediag}. This diagram illustrates the internship agreement workflow from the central internship unit’s perspective. The central unit can view all agreements that have been approved by the faculty. At this stage, the unit activates the agreement, and the official internship process starts. When the internship period ends, the central unit marks the agreement as \texttt{COMPLETED}. This stage ensures the formal completion of the internship process and corresponds to insurance termination in a real-life scenario.

Each component-based sequence diagram abstracts many internal validation steps. In this implementation, validation is divided between the backend and the chaincode. The backend checks authentication, input format, company availability, student metadata, and application-level business rules before sending a transaction. Then, the chaincode verifies the caller role, entity ownership, and allowed state transitions, and agreement-specific ledger rules. Therefore, the dynamic model is a combined backend-and-chaincode workflow rather than a chaincode-only process.

\subsubsection{Company Registration Request Lifecycle}

An additional dynamic process in the system is the company registration request flow. When a student wants to create an agreement with a company that is not already registered in the system, a request has to be submitted. The central internship unit reviews this request and decides whether to add the company to the system. If the request is approved, an identity and an account are created for the company so that it can log in to the system.

The figure below illustrates the company registration process:

\begin{figure}[H]
    \centering
    \includegraphics[width=1\linewidth]{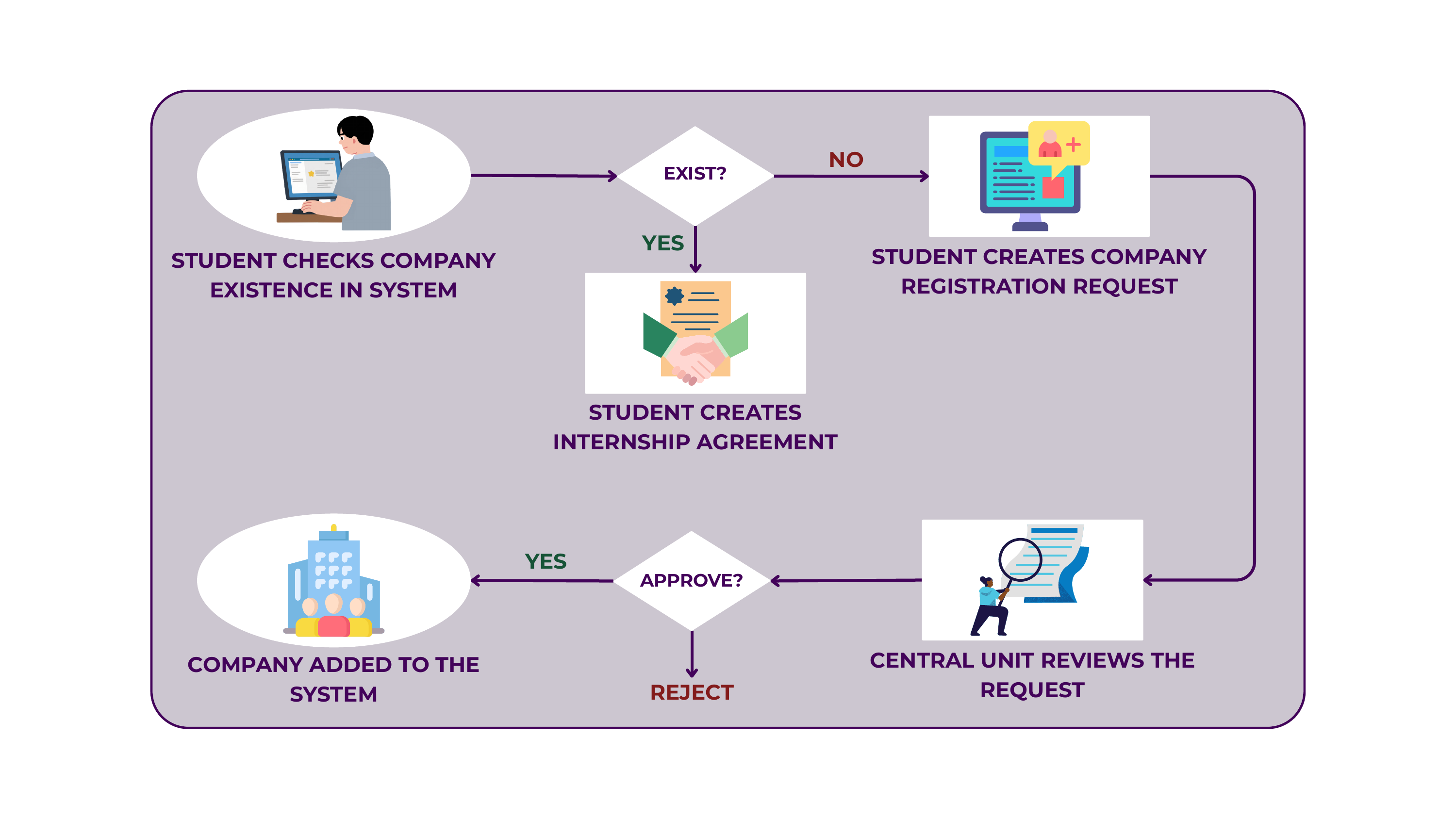}
    \caption{Company registration lifecycle model}
    \label{fig:companyregistrationflow}
\end{figure}

\newpage
\section{Experimentation Environment and Experiment Design}
\subsection{Development Requirements}
The implementation is developed and tested on a Linux environment running under WSL2. There is a blockchain layer with Hyperledger Fabric, a Node.js/Express backend, a JavaScript frontend, and an SQLite database for off-chain application data.

The blockchain layer is implemented with the Hyperledger Fabric test network. The smart contract, which is named `internship`, is developed in Node.js using the Fabric Contract API and deployed on the `mychannel` channel. The Fabric environment uses Docker containers for the peer, orderer, certificate authority, and chaincode runtime. The main tool versions are Node.js v18.20.8, npm 10.8.2, Docker 29.2.1, Docker Compose v5.0.2, Python 3.12.3, and Hyperledger Fabric peer v2.5.15. 

The backend uses the Fabric Gateway SDK to submit transactions to the chaincode and SQLite to store application-level metadata such as users, company registration information, and login credentials. The frontend communicates with the backend through REST API endpoints.

\subsection{Resource Requirements}
The system requires a machine to run Docker containers, Node.js services, and a local SQLite database. In this implementation, all Fabric identities are stored under the local identity directory. The main application settings, such as the channel name, chaincode name, MSP ID, CA URL, peer endpoint, and identity paths, are provided through environment variables. There are also default values for local use. 

\subsection{Functional Tests}
The correctness of the main workflows, access-control rules, and business rules is tested through functional tests. They were script-based tests and executed through the backend API. The tests include both positive scenarios and negative scenarios. Therefore, tests verify the successful completion of valid operations and the rejection of invalid operations. 

\begin{longtable}{p{4cm} p{9cm}}
\caption{Functional Test Coverage}
\label{tab:functional_tests} \\

\hline
\textbf{Test} & \textbf{Validated behavior} \\
\hline
\endfirsthead

\hline
\textbf{Test} & \textbf{Validated behavior} \\
\hline
\endhead

Workflow test & Verifies the complete internship agreement lifecycle. Includes creation by student, student approval, company approval, faculty approval, central activation, and completion. \\
\hline

Phase 2 workflow test & Verifies extended agreement creation features, including internship type, internship field, fixed working days, flexible weekly schedules, and agreement history. Also checks the rejection flow by creating an additional agreement, approving it as a student, and rejecting it as a company. \\
\hline

Rule enforcement test & Verifies business rules such as minimum completed credits, minimum internship start-date lead time, valid fixed or flexible weekly schedules, department-specific mandatory and voluntary internship constraints, overlap prevention, repeated mandatory field prevention, maximum mandatory internship count, and voluntary internship limits. \\
\hline

Company registration test & Verifies that a student can submit a company request and that the central unit can approve it, provision a Fabric identity, create company credentials, and checks that the generated company identity can successfully read data from the blockchain. \\
\hline

\end{longtable}

\subsubsection{Workflow Test}

The workflow test validates the full internship agreement lifecycle. Initially, a student creates an agreement, and the test checks whether the initial status is \texttt{CREATED}. Then, the test verifies the sequential status transitions from \texttt{CREATED} to \texttt{STUDENT\_APPROVED}, \texttt{COMPANY\_APPROVED}, \texttt{FACULTY\_APPROVED}, \texttt{ACTIVE}, and finally \texttt{COMPLETED}. 

Moreover, the test checks pending agreement lists for the company and faculty users to confirm that each actor sees the agreement at the correct stage. Also, the test validates access control and ledger history behavior. The central internship unit is expected to receive HTTP \texttt{403}, which is an error code for unauthorized access, when trying to read the agreement or its history before faculty approval. The central unit can read the agreement and history successfully only after faculty approval. 

\subsubsection{Phase 2 Workflow Test}

This test extends the main workflow by testing additional agreement fields and alternative branches. Different situations are tested using internship type, internship field, fixed working days, and flexible weekly schedules. Moreover, the test checks whether \texttt{totalWorkingDays} is calculated correctly and whether the submitted \texttt{weeklySchedule} structure is accepted.

Furthermore, this test checks the rejection flow by creating a new agreement, which is approved by the student and rejected by the company. The test verifies that the agreement status becomes \texttt{REJECTED} and that the rejecting role and state are consistent with the pre-defined rejection rules. Also, the rejected agreement history is queried to confirm that the rejection event is traceable.

\subsubsection{Rule Enforcement Test}
The rule enforcement test validates the business constraints applied before agreement creation. The test verifies whether mandatory internships are rejected when the student has not completed the minimum required credits. Furthermore, it checks whether the internship start date is less than 15 days away and whether the working days are within the allowable range.

Moreover, the test verifies department-specific internship rules such as the number of mandatory and voluntary internships, the required working-day limits, and the requirement to use different mandatory internship fields.

\subsubsection{Company Registration Test}

This test validates the controlled onboarding process for companies. Initially, the student submits a company registration request. Then, the central internship unit reads and approves the pending request. After approval, the system creates the company record, generates login credentials, produces a temporary password, and provisions a Fabric identity.

After the registration process ends, the generated company account logs into the system. Then, the company's agreement list is queried to check whether its Fabric identity is successfully created and whether the company can read data from the ledger.

Overall, these results confirm that the prototype enforces workflow order, role-based access control, business constraints, rejection handling, company onboarding, and ledger history tracking consistently across the backend and blockchain layers. 

\subsection{Performance Tests}
\label{perf_tests}
Performance tests were executed to measure the latency and throughput of the main read, ledger-write, and company-registration operations. The experiments were performed on the local Fabric test network. Thus, the results reflect a prototype-level performance rather than production-scale performance.

\subsubsection{Read-Only API and Ledger Query Performance}
The read-only benchmark tested agreement queries and approved-company listing. Agreement queries read data from Fabric chaincode with \texttt{evaluateTransaction} and the approved-company list reads from SQLite through the backend API.

For this benchmark, each read scenario was executed for 50 seconds after a 5-second warm-up period, with 10 concurrent workers sending requests to the backend API. The table reports the total number of completed requests, throughput in requests per second, average latency, and P95 latency for each scenario. P95 latency means that 95\% of requests finished within this response time.

The read-only performance test includes the following scenarios:

\begin{itemize}
    \item \texttt{student\_my\_agreements}: shows the internship agreements that belong to the student
    \item \texttt{student\_pending\_agreements}: shows the agreements that are waiting for student action
    \item \texttt{company\_pending\_agreements}: shows the agreements that are waiting for company approval
    \item \texttt{faculty\_pending\_agreements}: shows the agreements that are waiting for faculty approval
    \item \texttt{central\_my\_agreements}: shows the agreements that can be viewed by the central internship unit
    \item \texttt{approved\_companies}: shows the approved companies stored in the system
\end{itemize}

\begin{longtable}{lrrrr}
\caption{Read-Only Performance Results}
\label{tab:read_perf} \\

\hline
\textbf{Scenario} & \textbf{Requests} & \textbf{TPS} & \textbf{Avg. ms} & \textbf{P95 ms} \\
\hline
\endfirsthead

\hline
\textbf{Scenario} & \textbf{Requests} & \textbf{TPS} & \textbf{Avg. ms} & \textbf{P95 ms} \\
\hline
\endhead

student\_my\_agreements & 7983 & 159.53 & 62.66 & 76.47 \\
student\_pending\_agreements & 7685 & 153.60 & 65.08 & 79.79 \\
company\_pending\_agreements & 7054 & 140.96 & 70.92 & 89.22 \\
faculty\_pending\_agreements & 7029 & 140.46 & 71.16 & 90.85 \\
central\_my\_agreements & 6229 & 124.52 & 80.29 & 97.40 \\
approved\_companies & 45861 & 917.13 & 10.90 & 14.49 \\

\hline
\end{longtable}

The SQLite approved\_companies query achieved higher throughput and lower latency than Fabric agreement queries, which is expected because it does not require chaincode evaluation. 

The figures below show the Fabric read throughput per second (TPS) and latency:

\begin{figure}[H]
    \centering
    \includegraphics[width=1\linewidth]{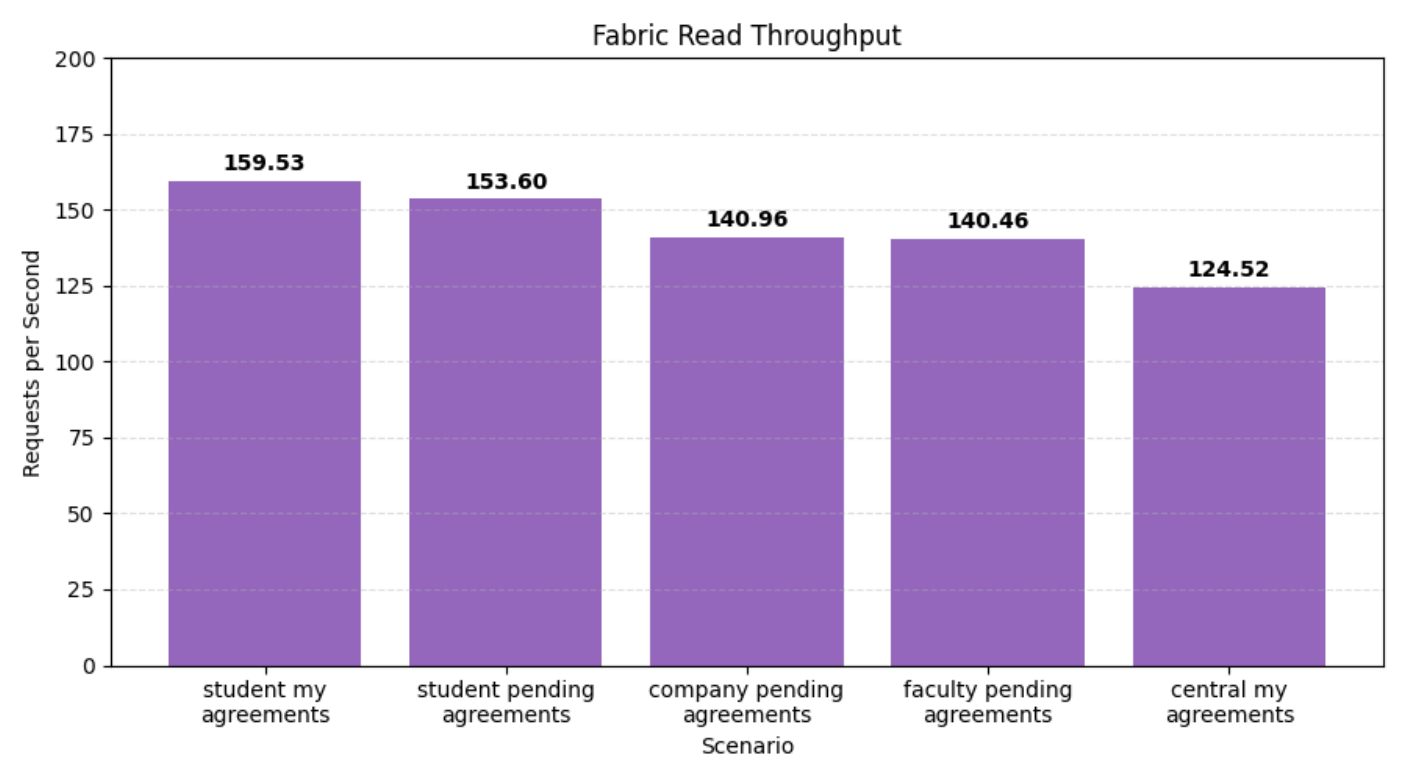}
    \caption{Fabric-Based Agreement Read Throughput}
    \label{fig:read-tps}
\end{figure}

\begin{figure}[H]
    \centering
    \includegraphics[width=1\linewidth]{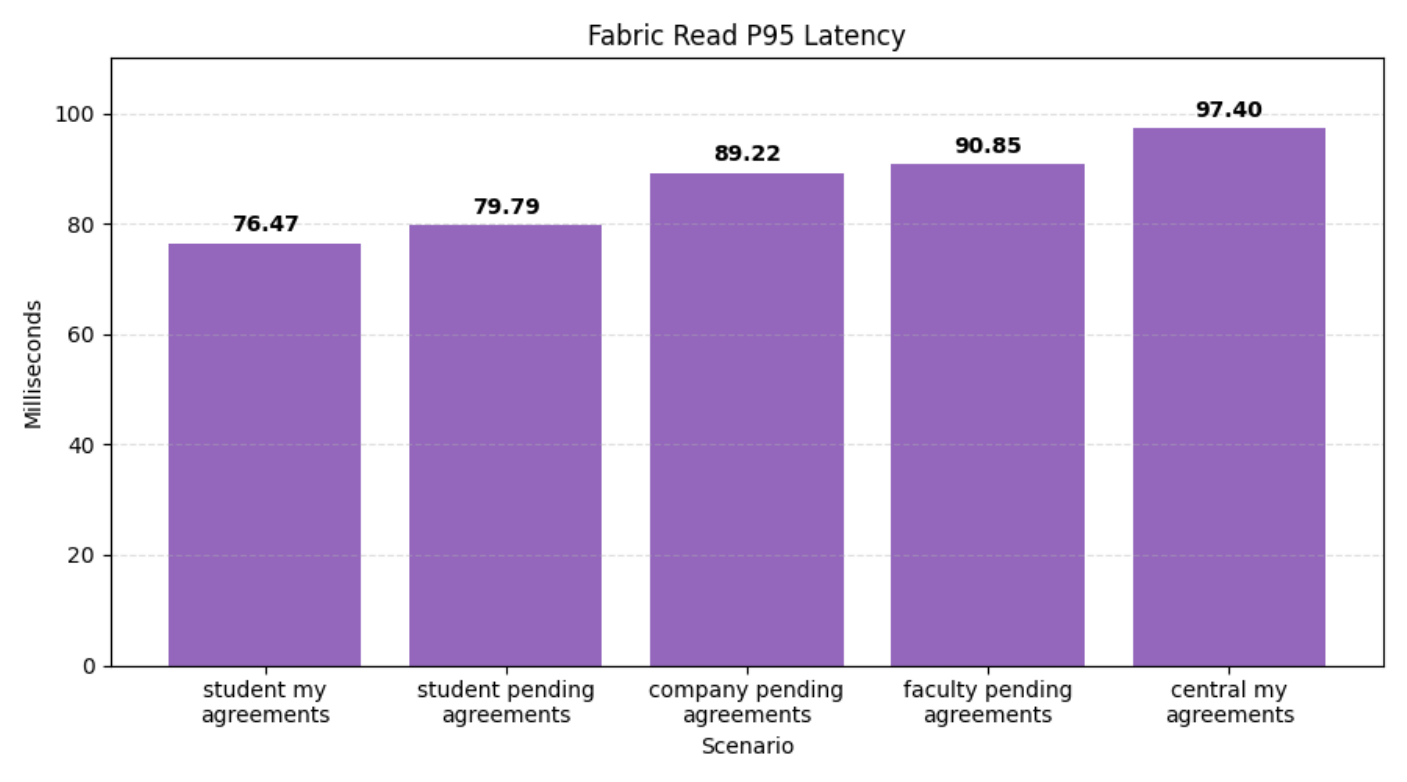}
    \caption{Fabric-Based Agreement Read Latency}
    \label{fig:read-latency}
\end{figure}

The figures below show SQLite/API read throughput per second (TPS) and latency:

\begin{figure}[H]
    \centering

    \begin{subfigure}{0.48\linewidth}
        \centering
        \includegraphics[width=\linewidth]{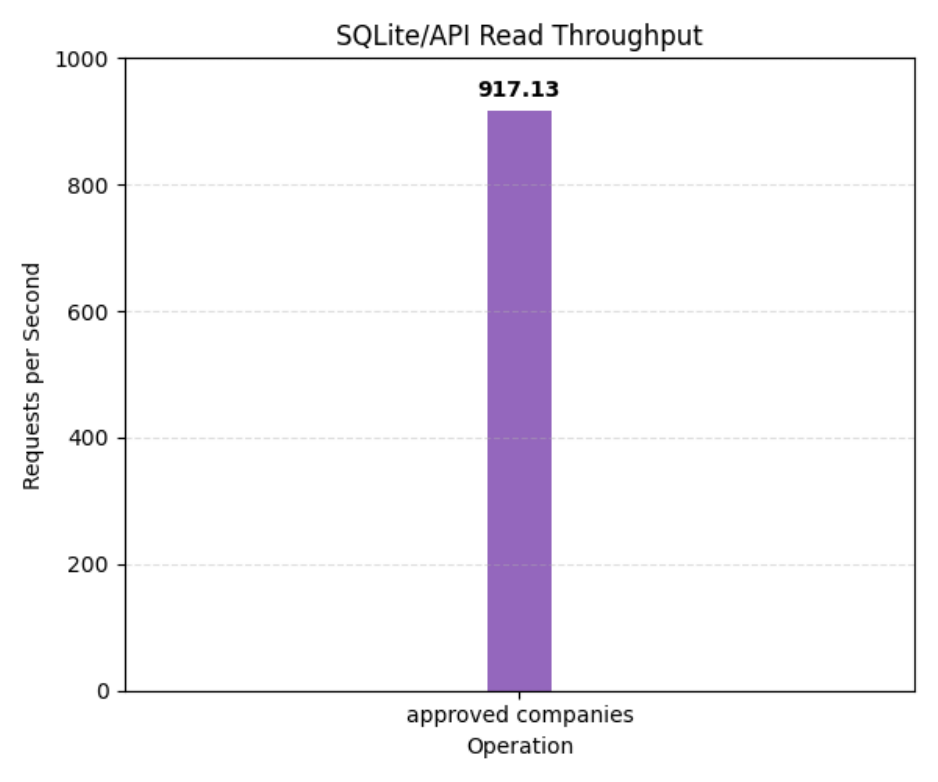}
        \caption{SQLite/API Read Throughput}
        \label{fig:sqlite-tps}
    \end{subfigure}
    \hfill
    \begin{subfigure}{0.48\linewidth}
        \centering
        \includegraphics[width=\linewidth]{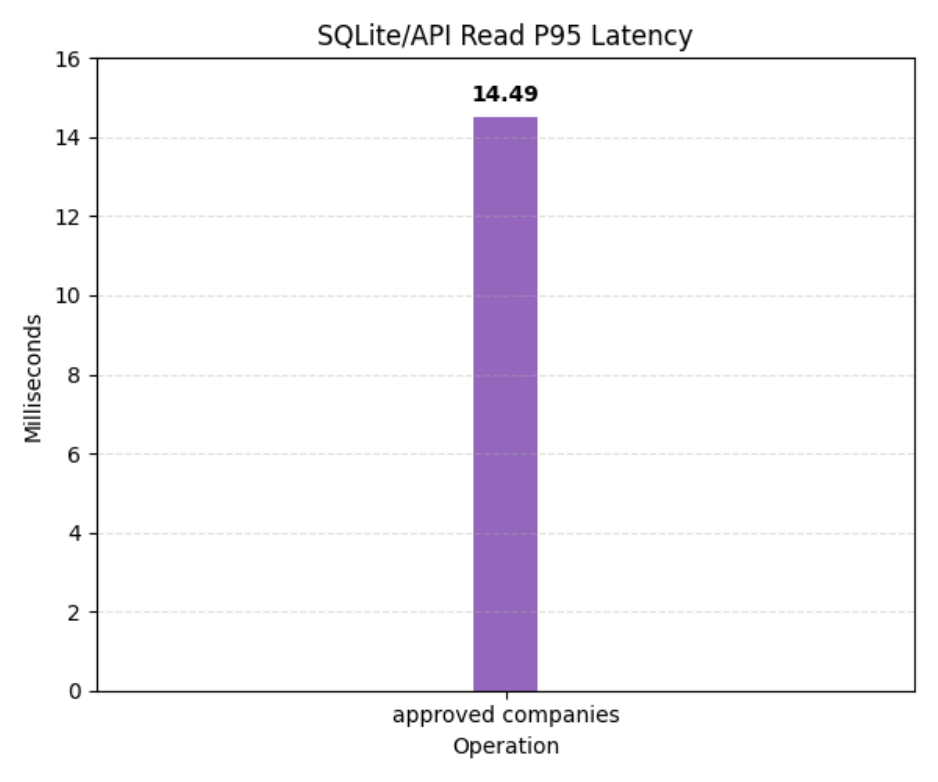}
        \caption{SQLite/API Read Latency}
        \label{fig:sqlite-latency}
    \end{subfigure}

    \caption{SQLite/API Read Performance Results}
    \label{fig:sqlite-read-performance}
\end{figure}

\subsubsection{Ledger Write Workflow Performance}
The ledger workflow benchmark measured a full agreement lifecycle with one iteration. This test includes all major approval stages, including student, company, and faculty approvals, and central-unit operations.

The ledger workflow performance test includes the following scenarios:

\begin{itemize}
    \item \texttt{create\_agreement}: creates a new internship agreement
    \item \texttt{student\_approve}: records the student approval
    \item \texttt{company\_approve}: records the company approval
    \item \texttt{faculty\_approve}: records the faculty approval
    \item \texttt{central\_activate}: activates the agreement by the central internship unit
    \item \texttt{central\_complete}: completes the agreement by the central internship unit
    \item \texttt{read\_faculty\_approved\_agreement}: reads the agreement after faculty approval
    \item \texttt{read\_agreement\_history}: reads the previous records of the agreement from the ledger
\end{itemize}

\begin{longtable}{lrrrr}
\caption{Ledger Full Workflow Performance Results}
\label{tab:ledger_perf} \\

\hline
\textbf{Operation} & \textbf{Type} & \textbf{Avg.ms} & \textbf{P95 ms} \\
\hline
\endfirsthead

\hline
\textbf{Operation} & \textbf{Type} & \textbf{Avg.ms} & \textbf{P95 ms} \\
\hline
\endhead

create\_agreement & write & 2117.28 & 2117.28 \\
student\_approve & write & 2047.57 & 2047.57 \\
company\_approve & write & 2059.85 & 2059.85 \\
faculty\_approve & write & 2043.63 & 2043.63 \\
central\_activate & write & 2037.67 & 2037.67 \\
central\_complete & write & 2039.05 & 2039.05 \\
read\_faculty\_approved\_agreement & read & 22.11 & 22.11 \\
read\_agreement\_history & read & 32.19 & 32.19 \\

\hline
\end{longtable}

Each ledger write operation, such as create\_agreement, student\_approve, company\_approve, faculty\_approve, central\_activate, and central\_complete, took approximately two seconds. This latency includes backend validation, Fabric communication, transaction processing, response parsing, and saving local metadata when needed. On the other hand, read operations such as read\_faculty\_approved\_agreement and read\_agreement\_history were faster and remained below 35 ms. 

The figures below show the Fabric read and write transaction throughput per second (TPS) and latency: 

\begin{figure}[H]
    \centering
    \includegraphics[width=1\linewidth]{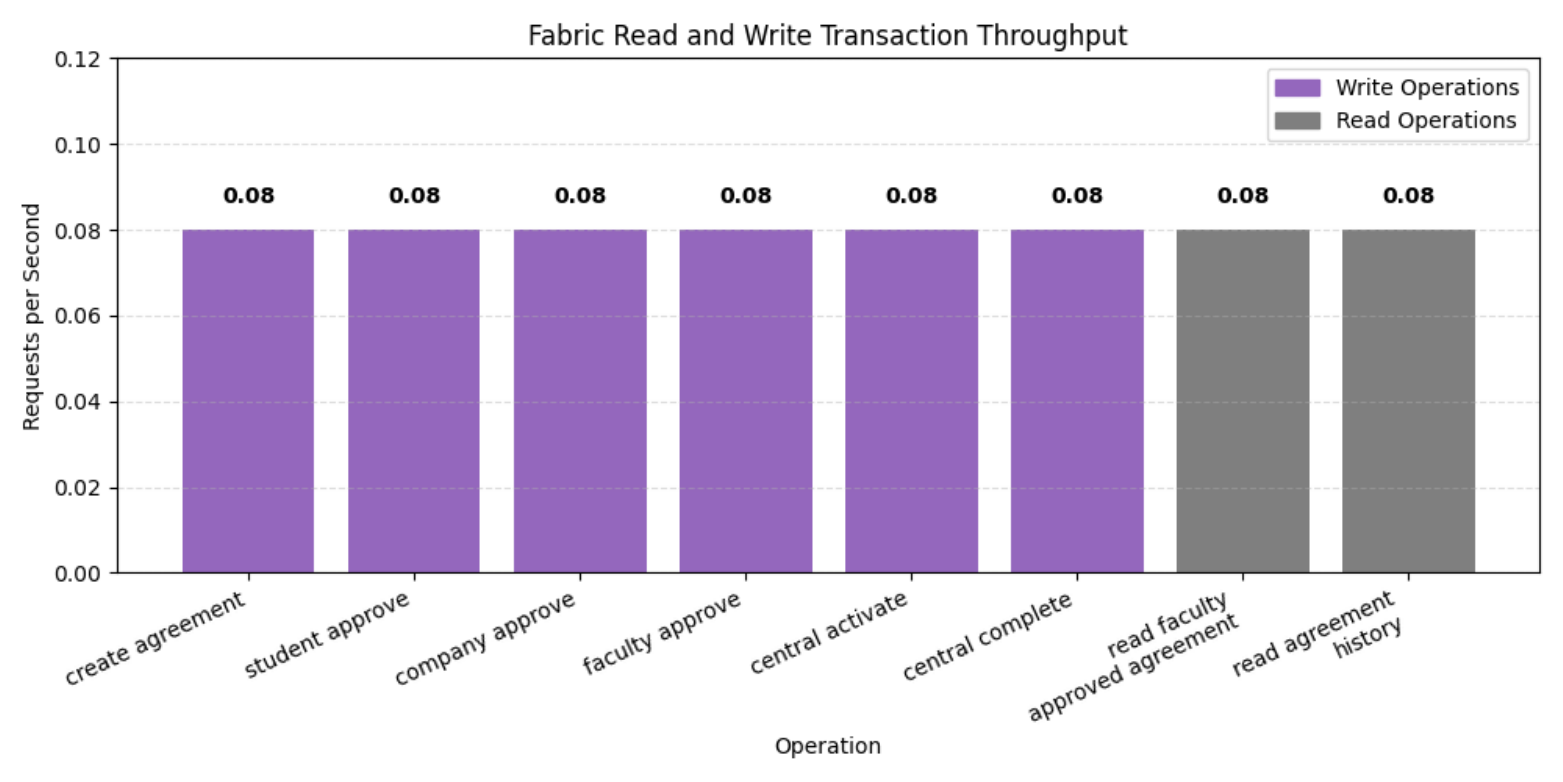}
    \caption{Fabric-Based Read and Write Transaction Throughput}
    \label{fig:rw-tps}
\end{figure}

\begin{figure}[H]
    \centering
    \includegraphics[width=1\linewidth]{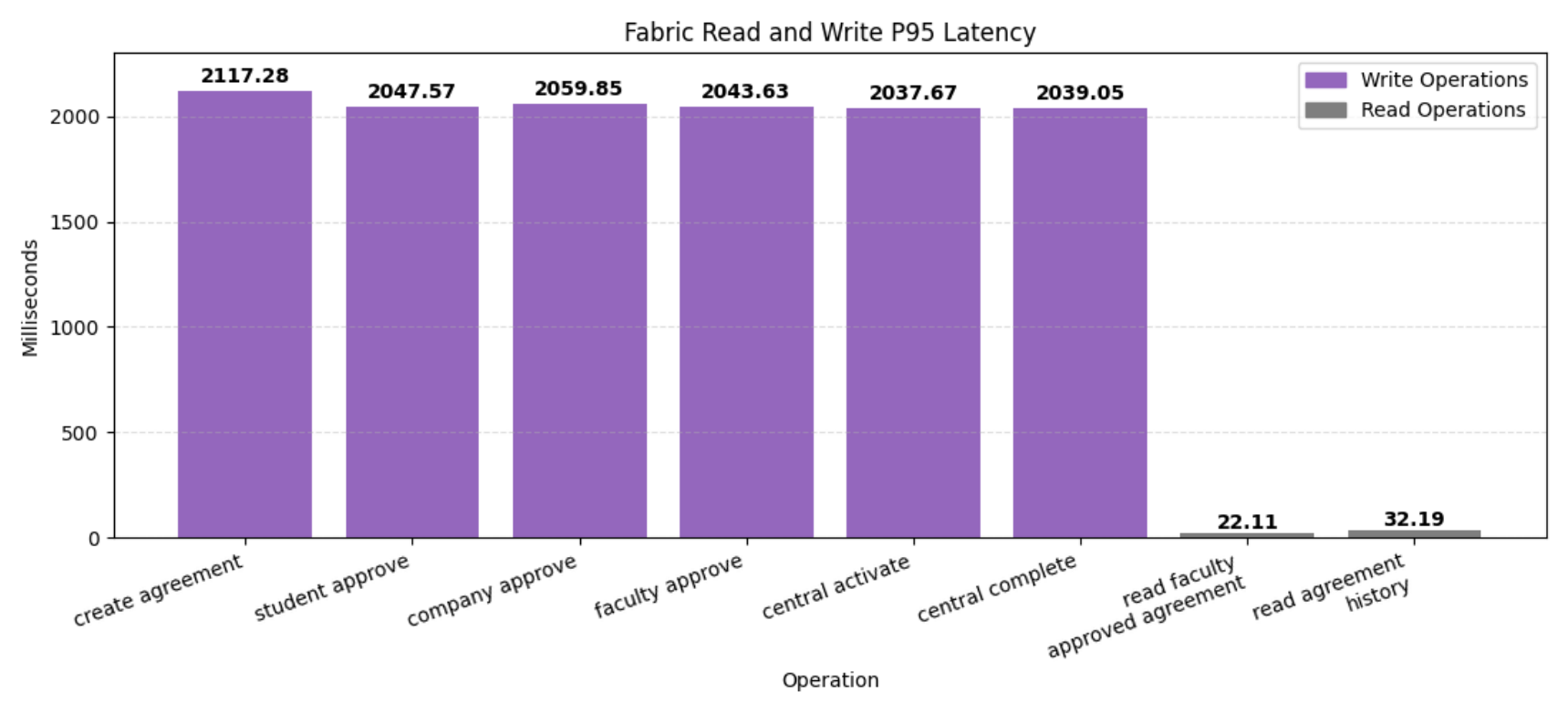}
    \caption{Fabric-Based Read and Write Transaction Latency}
    \label{fig:rw-latency}
\end{figure}

As an important note, the TPS values for the full internship agreement workflow are lower since this benchmark executes a sequential lifecycle scenario rather than a continuous multi-worker load test. Thus, these values should be interpreted as workflow execution results, not as maximum Fabric throughput.

\subsubsection{Company Registration Performance}
The company registration benchmark tested the company onboarding process and Fabric identity creation. The benchmark was repeated 20 times.

The company registration performance test includes the following scenarios:

\begin{itemize}
    \item \texttt{submit\_company\_request}: submits a new company request by a student
    \item \texttt{list\_pending\_company\_requests}: lists the company requests waiting for review
    \item \texttt{read\_company\_request\_detail}: reads the details of a company request
    \item \texttt{approve\_company\_request}: approves the company request and creates the company record and identity
    \item \texttt{login\_created\_company}: logs in with the newly created company account
    \item \texttt{read\_created\_company\_agreements}: reads agreement data with the newly created company account
\end{itemize}

\begin{longtable}{lrrrr}
\caption{Company Registration Performance Results}
\label{tab:company_perf} \\

\hline
\textbf{Operation} & \textbf{Requests} & \textbf{Avg. ms} & \textbf{P95 ms} \\
\hline
\endfirsthead

\hline
\textbf{Operation} & \textbf{Requests} & \textbf{Avg. ms} & \textbf{P95 ms} \\
\hline
\endhead

submit\_company\_request & 20 & 12.42 & 18.92 \\
list\_pending\_company\_requests & 20 & 2.99 & 4.02 \\
read\_company\_request\_detail & 20 & 2.90 & 3.87 \\
approve\_company\_request & 20 & 370.03 & 397.99 \\
login\_created\_company & 20 & 61.62 & 65.47 \\
read\_created\_company\_agreements & 20 & 20.74 & 22.48 \\

\hline
\end{longtable}

\begin{figure}[H]
    \centering
    \includegraphics[width=1\linewidth]{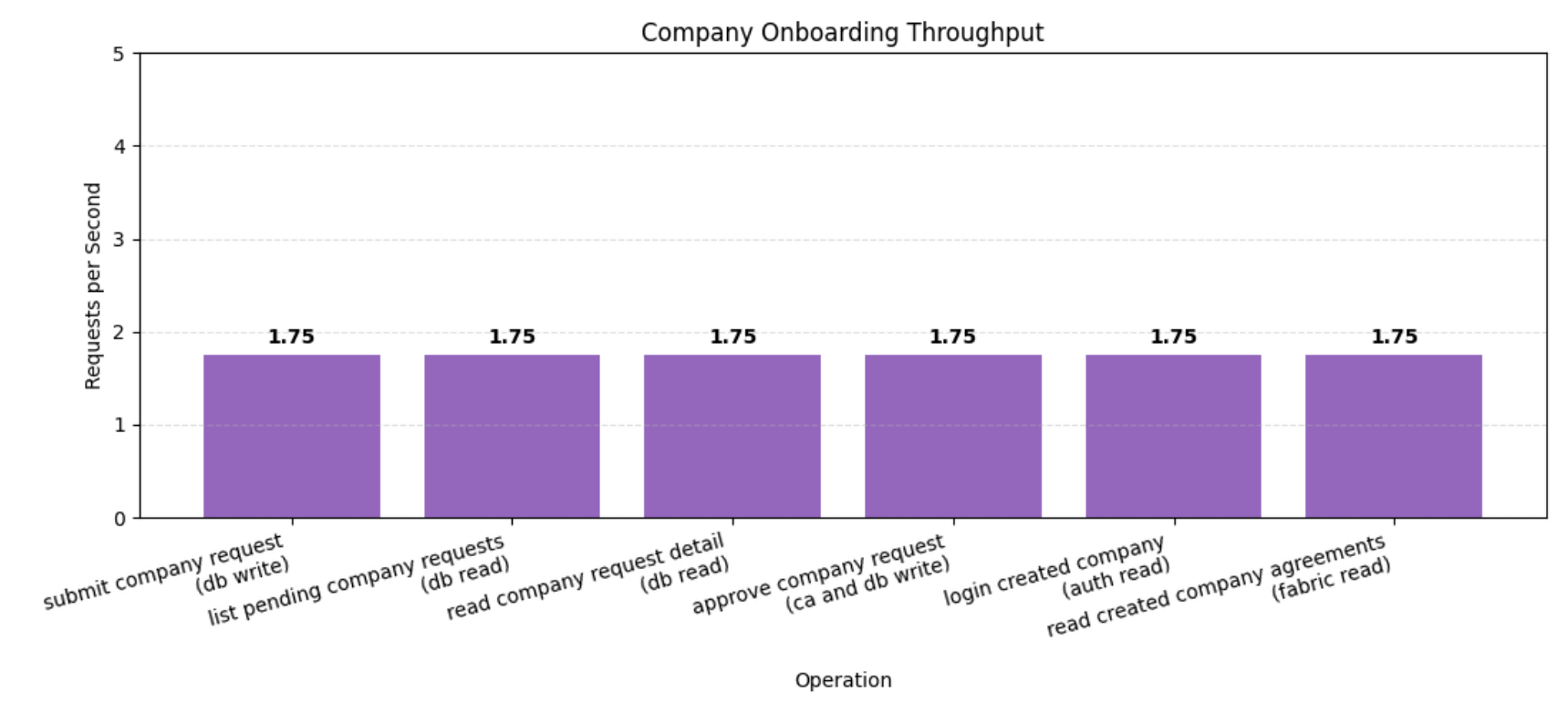}
    \caption{Company Onboarding Transaction Throughput}
    \label{fig:comp-tps}
\end{figure}

\begin{figure}[H]
    \centering
    \includegraphics[width=1\linewidth]{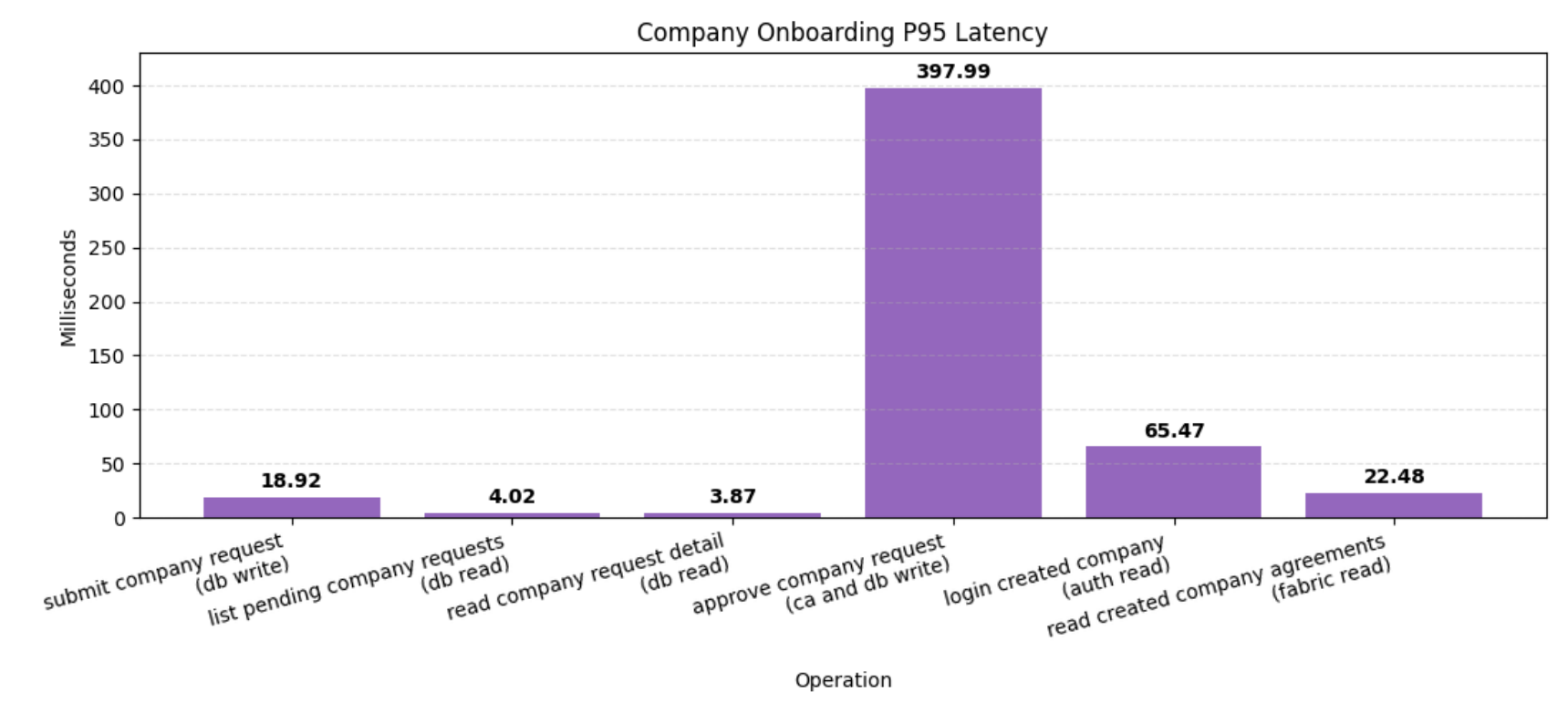}
    \caption{Company Onboarding Transaction Latency}
    \label{fig:comp-latency}
\end{figure}

The approval step had the highest latency because it includes Fabric CA identity provisioning, SQLite writes, password hashing, and user-account creation.

\newpage
\section{Comparative Evaluation and Discussion}
\label{comp_eval}

The StajChain project is evaluated in terms of functional coverage, performance results, and comparison with existing Hyperledger Fabric performance studies. In addition, the criteria and goals defined in the interim report will be used to assess the success of the project. 

\subsection{Supported Functional Features}
\label{supported_features}

Before presenting the performance results, it is useful to outline the functionality of the implemented prototype. This part shows which system features were implemented and how the main design goals, such as role-based access control, workflow management, traceability, and data integrity, are supported.

The table below summarizes the main functional features supported by the StajChain prototype:

\begin{table}[H]
\centering
\small
\setlength{\tabcolsep}{4pt}
\renewcommand{\arraystretch}{1.15}
\caption{Supported Functional Features of StajChain}
\label{tab:supported_features}
\begin{tabularx}{1\textwidth}{|p{4cm}|X|}
\hline
\textbf{Feature} & \textbf{Description} \\ \hline

Permissioned access & 
Only registered and authorized actors can participate in the system. \\ \hline

Role-based access control & 
Students, companies, faculty users, and the central unit have different permissions. \\ \hline

Agreement lifecycle management & 
The system manages agreement creation, approval, activation, completion, and rejection. \\ \hline

Rejection handling & 
Authorized actors can reject agreements only at allowed workflow stages. \\ \hline

Sequential workflow control & 
Agreement actions must follow the predefined student, company, faculty, and central unit order. \\ \hline

On-chain state transition control & 
Smart contracts control agreement status changes and prevent invalid transitions. \\ \hline

Blockchain-based integrity & 
Critical agreement records and state changes are stored on the ledger. \\ \hline

Transparent agreement history & 
Previous agreement actions can be queried through ledger history. \\ \hline

Company onboarding & 
Companies are reviewed and approved before they become participants in the system. \\ \hline

Fabric identity provisioning & 
Approved companies are assigned Fabric identities during onboarding. \\ \hline

Internship-specific rule enforcement & 
The system checks internship type, field, working days, total days, credit requirements, and overlap rules. \\ \hline

Off-chain metadata management & 
Application-level data such as users, companies, requests, and scheduling metadata are stored off-chain. \\ \hline

\end{tabularx}
\end{table}

These features show that StajChain is not only a record storage system. It also combines permissioned access, workflow control, company onboarding, internship-specific validation, and ledger-based traceability in the same prototype.

\subsection{Performance Comparison with Existing Fabric Performance Studies}

Several studies in the literature evaluate Hyperledger Fabric using quantitative measurements such as throughput, latency, execution time, scalability, and resource usage. The StajChain project follows the same evaluation approach by measuring throughput and P95 latency for main operations. However, direct comparison is not possible because each study uses different Fabric versions, network topologies, hardware resources, endorsement policies, and transaction types. Therefore, these studies are used to provide a quantitative performance context rather than a direct benchmark ranking. Among these studies, Dash et al. is especially useful for comparison since it focuses on an academic institutional process. This study also compares Hyperledger Fabric with public blockchain platforms using TPS and latency values.

The table below compares StajChain with different studies according to their performance evaluations:

\begin{table}[H]
\centering
\scriptsize
\setlength{\tabcolsep}{3pt}
\renewcommand{\arraystretch}{1.05}
\caption{Quantitative Comparison with Existing Fabric Studies}
\label{tab:literature_quantitative_comparison}
\begin{tabularx}{1\textwidth}{|p{2.4cm}|p{2.2cm}|p{3.2cm}|X|}
\hline
\textbf{Study} & \textbf{Metric(s)} & \textbf{Reported Result} & \textbf{Relation to StajChain} \\ \hline

Nasir et al. \cite{nasir2018performance} &
Execution time, latency, throughput, scalability &
Fabric v1.0 reached up to 490 TPS for queries (read-like operations) and 185 TPS for invoke (write-like operations). &
StajChain results also show that read operations are faster than ledger write operations. \\ \hline

Sukhwani et al. \cite{sukhwani2018modeling} &
Throughput, utilization, queue length, latency &
Endorsement, ordering, validation, and ledger write stages affect Fabric performance. &
This supports the higher latency of StajChain write operations. \\ \hline

Thakkar et al. \cite{thakkar2018fabric} &
Throughput and latency &
Throughput of the initial Fabric setup reached around 140 TPS. After optimizations, throughput reached 2250 TPS. &
StajChain read throughput is 124.52--159.53 TPS, which is close to the initial setup range. \\ \hline

Guggenberger et al. \cite{guggenberger2022performance} &
Throughput, latency, scalability, database effects &
Fabric performance changes according to database type, workload, network size, and transaction type. &
This supports evaluating StajChain results through its own prototype setup and workload rather than direct ranking.  \\ \hline

Abbasi et al. \cite{abbasi2025benchmarking} &
Throughput, latency, send rate &
Throughput values of about 210.9 TPS, 350.8 TPS, and 605.5 TPS were reported for different network setups. &
This supports that setup and workload affect Fabric performance. \\ \hline

Dash et al. \cite{dash2025hyperledger} &
TPS and average latency &
HLF reached 130--280 TPS and 300--900 ms latency for academic certificate authentication. & 
StajChain read TPS is close to the lower values Dash et al. reports.  \\ \hline

\end{tabularx}
\end{table}

This comparison shows that throughput and latency are the main performance criteria used in Hyperledger Fabric studies. Moreover, the reported Hyperledger Fabric TPS values support the suitability of Hyperledger Fabric for academic systems that require verification, workflow management, and access control. The StajChain results are sufficient for the expected internship agreement workload.

\subsection{Quantitative Results Overview}

This section provides an overview of all quantitative results obtained in the performance evaluation. The detailed results are presented in Section \ref{perf_tests}. The two subsections below summarize all throughput and P95 latency results of the main operations.

\subsubsection{Overall Throughput Results}

\begin{table}[H]
\centering
\scriptsize
\setlength{\tabcolsep}{4pt}
\renewcommand{\arraystretch}{1.15}
\caption{Overall Throughput Results}
\label{tab:overall_tps_results}
\begin{tabularx}{0.95\textwidth}{|>{\raggedright\arraybackslash}X|p{2.3cm}|c|}
\hline
\textbf{Operation / Scenario} & \textbf{Group} & \textbf{TPS} \\ \hline

\multicolumn{3}{|c|}{\textbf{Read-Only Operations}} \\ \hline
Student agreements & Fabric read & 159.53 \\ \hline
Student pending agreements & Fabric read & 153.60 \\ \hline
Company pending agreements & Fabric read & 140.96 \\ \hline
Faculty pending agreements & Fabric read & 140.46 \\ \hline
Central agreements & Fabric read & 124.52 \\ \hline
Approved companies & SQLite read & 917.13 \\ \hline

\multicolumn{3}{|c|}{\textbf{Full Workflow Operations}} \\ \hline
Create agreement & Fabric write & 0.08 \\ \hline
Student approve & Fabric write & 0.08 \\ \hline
Company approve & Fabric write & 0.08 \\ \hline
Faculty approve & Fabric write & 0.08 \\ \hline
Central activate & Fabric write & 0.08 \\ \hline
Central complete & Fabric write & 0.08 \\ \hline
Read faculty-approved agreement & Fabric read & 0.08 \\ \hline
Read agreement history & Fabric read & 0.08 \\ \hline

\multicolumn{3}{|c|}{\textbf{Company Onboarding Operations}} \\ \hline
Submit company request & Onboarding & 1.75 \\ \hline
List pending company requests & Onboarding & 1.75 \\ \hline
Read company request detail & Onboarding & 1.75 \\ \hline
Approve company request & Onboarding & 1.75 \\ \hline
Login created company & Onboarding & 1.75 \\ \hline
Read created company agreements & Onboarding & 1.75 \\ \hline

\end{tabularx}
\end{table}

The throughput results show that read-only operations have much higher TPS than full workflow operations that include write operations. This is expected because read-only tests repeatedly execute query operations, while the full workflow test executes the agreement lifecycle step by step as a single scenario. Therefore, the workflow TPS values should not be interpreted as the highest TPS that a single operation can reach.

\subsubsection{Overall P95 Latency Results}

\begin{table}[H]
\centering
\scriptsize
\setlength{\tabcolsep}{4pt}
\renewcommand{\arraystretch}{1.15}
\caption{Overall P95 Latency Results}
\label{tab:overall_latency_results}
\begin{tabularx}{0.95\textwidth}{|>{\raggedright\arraybackslash}X|p{2.3cm}|c|}
\hline
\textbf{Operation / Scenario} & \textbf{Group} & \textbf{P95 ms} \\ \hline

\multicolumn{3}{|c|}{\textbf{Read-Only Operations}} \\ \hline
Student my agreements & Fabric read & 76.47 \\ \hline
Student pending agreements & Fabric read & 79.79 \\ \hline
Company pending agreements & Fabric read & 89.22 \\ \hline
Faculty pending agreements & Fabric read & 90.85 \\ \hline
Central my agreements & Fabric read & 97.40 \\ \hline
Approved companies & SQLite read & 14.49 \\ \hline

\multicolumn{3}{|c|}{\textbf{Full Workflow Operations}} \\ \hline
Create agreement & Fabric write & 2117.28 \\ \hline
Student approve & Fabric write & 2047.57 \\ \hline
Company approve & Fabric write & 2059.85 \\ \hline
Faculty approve & Fabric write & 2043.63 \\ \hline
Central activate & Fabric write & 2037.67 \\ \hline
Central complete & Fabric write & 2039.05 \\ \hline
Read faculty-approved agreement & Fabric read & 22.11 \\ \hline
Read agreement history & Fabric read & 32.19 \\ \hline

\multicolumn{3}{|c|}{\textbf{Company Onboarding Operations}} \\ \hline
Submit company request & Onboarding & 18.92 \\ \hline
List pending company requests & Onboarding & 4.02 \\ \hline
Read company request detail & Onboarding & 3.87 \\ \hline
Approve company request & Onboarding & 397.99 \\ \hline
Login created company & Onboarding & 65.47 \\ \hline
Read created company agreements & Onboarding & 22.48 \\ \hline

\end{tabularx}
\end{table}

The P95 latency results also show the difference between read and write operations. Read operations are below 100 ms P95 latency, while SQLite reads were faster with 14.49 ms P95 latency. On the other hand, write operations had approximately 2 seconds of P95 latency because they update the ledger completely. Moreover, company onboarding operations were mostly fast except the company approval operation. This is expected since it creates the company record, user account, and Fabric identity.

\subsection{Comparison of StajChain Against Interim Evaluation Criteria}

This section compares the implemented prototype with the evaluation criteria defined in the interim report.

The table below compares the goals defined in the report and the StajChain results:

\begin{table}[H]
\centering
\scriptsize
\setlength{\tabcolsep}{4pt}
\renewcommand{\arraystretch}{1.15}
\caption{Assessment Against Interim Evaluation Criteria}
\label{tab:interim_criteria_assessment}
\begin{tabularx}{1\textwidth}{|p{3.4cm}|p{3.5cm}|X|}
\hline
\textbf{Interim Evaluation Criterion} & \textbf{Goal} & \textbf{StajChain Result} \\ \hline

Functional correctness &
All agreement lifecycle stages should execute correctly without data inconsistency. &
Achieved. The functional tests showed that creation, approval, activation, completion, rejection, company onboarding, and invalid cases work as expected. \\ \hline

Transaction latency &
Blockchain transactions should be completed within 5 seconds under normal conditions. &
Achieved. Main Fabric write operations had about 2 seconds P95 latency. \\ \hline

System throughput &
The system should be able to process at least 20-30 transactions per minute. &
Partially achieved. Read operations achieved much higher throughput. However, the full workflow TPS values are lower because they measure a complete lifecycle scenario, not the maximum TPS of a single operation. Thus, the throughput target is reasonable for normal administrative use, but it should be tested with a separate write throughput test. \\ \hline

Query response time &
Read-only query operations should be completed within 1 second. &
Achieved. Fabric read operations stayed below 100 ms P95 latency, and the SQLite read operation had 14.49 ms P95 latency. \\ \hline

Ledger consistency &
Ledger state should remain consistent across all peers after transaction commitment. &
Achieved. The functional tests show that agreement states and history records returned the expected data after transactions. \\ \hline

Success rate &
At least 99\% of submitted transactions should be successfully committed to the ledger under normal workload conditions. &
Achieved in the measured benchmark scenarios. The recorded API operations in the read, ledger, and company-registration benchmarks completed with 100\% success rate, while invalid functional-test cases were rejected as expected. \\ \hline

Access control &
Role-based access control enforced by smart contracts should be verified through testing. &
Achieved. Unauthorized actions and incorrect workflow transitions returned the expected errors. \\ \hline

Data integrity &
Internship agreement records stored on the ledger must remain immutable. Any modification should not overwrite existing records and should result in new ledger entries. &
Achieved. Agreement updates are stored as state transitions, and previous actions can be viewed using ledger history. \\ \hline

Usability &
Users with no technical background should be able to submit and verify internship agreements using the web interface. &
Partially achieved. The prototype provides a basic web interface for the main actors. However, it is important to note that a formal usability test was not conducted with end users. \\ \hline

\end{tabularx}
\end{table}

\subsection{Discussion}
\label{discussion}

The system provides a digital internship agreement process with role-based access control, workflow management, company onboarding, and ledger history. The implemented prototype supports agreement creation, approval, activation, completion, rejection, and history queries. Moreover, the test results indicate that the StajChain prototype satisfies the main goals of the project. 

The functional test results show that the prototype operated as intended. As shown in Table~\ref{tab:functional_tests}, the tests covered the complete agreement lifecycle, agreement creation rules, business-rule enforcement, and company registration workflow. Other test cases, such as invalid inputs, unauthorized actions, and incorrect workflow transitions, returned the expected error responses. As a result, the functional test results indicate that StajChain enforces role-based access control, agreement state transitions, company onboarding, and internship-specific business rules correctly.

Furthermore, the performance test results indicate that the StajChain project is suitable for administrative internship workflows. However, StajChain is a system for the internship agreement process and not a general benchmark to measure the maximum performance of Hyperledger Fabric. Therefore, the results should be interpreted according to this context.

The test results show that read operations are much faster than ledger write operations. As shown in Table~\ref{tab:overall_tps_results}, Fabric-based agreement read operations achieved between 124.52 and 159.53 TPS with P95 latency below 100 ms. On the other hand, the SQLite-based approved companies query reached 917.13 TPS with 14.49 ms P95 latency. This is expected because SQLite reads do not require chaincode execution or ledger transaction processing. 

Fabric write operations were slower than read operations. Main ledger write operations had approximately 2 seconds P95 latency, which is below the 5-second transaction latency target in the interim report. This result is expected because Fabric write transactions require endorsement, ordering, validation, and commit processing before updating the ledger state. 

Moreover, throughput results are in the expected range. The read throughput results in Table~\ref{tab:overall_tps_results} satisfied the 20--30 transactions per minute target of the interim report. However, the full workflow TPS values are lower because they measure a complete lifecycle scenario, not the maximum TPS of a single operation.

Lastly, the company registration results in Table~\ref{tab:overall_latency_results} show that all onboarding operations except company approval had low P95 latency. The company approval operation had 397.99 ms P95 latency, which is expected since this step includes Fabric identity provisioning, database writes, and user account creation. Since the company approval operation is not frequent, this latency is acceptable for the system. 

Overall, StajChain meets the main functional and performance goals defined in the interim report. It uses Hyperledger Fabric for controlled multi-party workflow execution, traceability, and tamper-resistant agreement history. The performance results are also at an acceptable level for the expected system usage. Thus, StajChain is suitable for administrative internship agreement workflows where correctness, traceability, auditability, and controlled access are more important than high-frequency transaction processing.

\newpage
\section{Conclusion and Future Work}
\label{conclusion}
The StajChain project is a permissioned blockchain-based system implemented to manage multi-party internship agreement processes. The main motivation of this project was the dependency on paper-based documents, handwritten signatures, and manual verification steps in the internship approval process. StajChain digitalizes the internship agreement lifecycle by using Hyperledger Fabric. In this project, each actor has a different role, and each agreement follows a predefined sequence of steps. The main actors in this project are students, companies, faculty users, and the central internship unit. Also, the system enforces role-based access control, internship constraints, company registration, agreement approval, rejection, activation, completion, and ledger history queries. 

Multiple tests were conducted to evaluate whether the system operates as intended and to measure its performance. The functional tests confirmed the main workflows, access-control rules, and business rules. Meanwhile, the performance tests indicated that read-only operations have low latency, while ledger write operations require higher latency. This higher latency occurs because each ledger write operation must be validated, ordered, and committed by Hyperledger Fabric. These results show that the proposed design is suitable for administrative internship workflows, where correctness, traceability, and auditability are more important than very high transaction throughput.

As future work, the prototype can be deployed on a distributed multi-organization Hyperledger Fabric network. In such a deployment, the university, companies, and other institutional actors can operate as separate organizations. This would make the system closer to a real deployment environment. Another possible improvement is the integration of secure document storage, such as IPFS. In the current prototype, internship-related application data is stored in SQLite. However, future versions may store document files off-chain in a distributed storage layer and keep their cryptographic hashes on the ledger for integrity verification. Moreover, additional verification steps can be added to the company registration process, such as checking tax identification numbers using governmental systems. Lastly, notification mechanisms and simple reporting tools can be added to make the system more useful for institutions.

\newpage
\bibliographystyle{IEEEtran}
\bibliography{references.bib} 

\end{document}